# METAHEURISTIC APPROACHES TO REALISTIC PORTFOLIO OPTIMISATION

by

**FRANCO RAOUL BUSETTI**

submitted in part fulfillment of the requirements

for the degree of

**MASTER OF SCIENCE**

in the subject

**OPERATIONS RESEARCH**

at the

**UNIVERSITY OF SOUTH AFRICA**

Supervisor: **PROF P H POTGIETER**

**June 2000**

# Abstract

In this thesis we investigate the application of two heuristic methods, genetic algorithms and tabu/scatter search, to the optimisation of realistic portfolios. The model is based on the classical mean-variance approach, but enhanced with floor and ceiling constraints, cardinality constraints and nonlinear transaction costs which include a substantial illiquidity premium, and is then applied to a large 100-stock portfolio.

It is shown that genetic algorithms can optimise such portfolios effectively and within reasonable times, without extensive tailoring or fine-tuning of the algorithm. This approach is also flexible in not relying on any assumed or restrictive properties of the model and can easily cope with extensive modifications such as the addition of complex new constraints, discontinuous variables and changes in the objective function.

The results indicate that that both floor and ceiling constraints have a substantial negative impact on portfolio performance and their necessity should be examined critically relative to their associated administration and monitoring costs.

Another insight is that nonlinear transaction costs which are comparable in magnitude to forecast returns will tend to diversify portfolios; the effect of these costs on portfolio risk is, however, ambiguous, depending on the degree of diversification required for cost reduction. Generally, the number of assets in a portfolio invariably increases as a result of constraints, costs and their combination.

The implementation of cardinality constraints is essential for finding the best-performing portfolio. The ability of the heuristic method to deal with cardinality constraints is one of its most powerful features.

*Keywords*: portfolio optimisation, efficient frontier, heuristic, genetic algorithm, tabu search

I declare that

**METAHEURISTIC APPROACHES TO REALISTIC PORTFOLIO OPTIMISATION**

is my own work and that all the sources that I have used or quoted have been indicated and acknowledged by means of complete references.

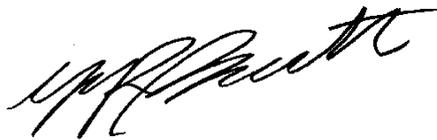

Franco Busetti

460-431-8

**To Barbara**

**for the self-heuristics**[1]

---

[1] Etymologically the word *heuristic* comes from the Greek word *heuriskein,* to discover; the Greek mathematician and inventor Archimedes (287-212 B.C.) is known for the famous *Heureka!* when he discovered a method for determining the purity of gold.

# Table of contents

# List of tables

# List of figures

# 1. Introduction

## 1.1 Background

A core function of the fund management industry is the combination of assets that appear attractive on a stand-alone basis into portfolios. These portfolios are required to be optimal in the sense of balancing the conflicting aspects of returns and risk. While the basis for portfolio optimisation was established by Markowitz [1] in a seminal article almost 50 years ago, it is often difficult to incorporate real-world constraints and dilemmas into the classical theory, which can limit its use. Although quantitative approaches to portfolio optimisation are becoming more widely adopted, the major portion of portfolio selection decisions continues ultimately to be taken on a qualitative basis.

Markowitz' mean-variance model of portfolio selection is one of the best-known models in finance and was the bedrock of modern portfolio theory. However, it is simplistic in that some of the underlying assumptions are not met in practice and it also ignores practical considerations such as transaction costs, liquidity constraints (and the resulting nonlinearities in transaction costs which result from this), minimum lot sizes and cardinality constraints, i.e. the restriction of a portfolio to a certain number of assets.

Incorporating all these considerations in the model results in a nonlinear mixed-integer programming problem which is substantially more difficult to solve. Exact solutions are unsuccessful when applied to large-scale problems and the approximations introduced to make these soluble are often unrealistically simplistic.

While large commercial portfolio optimisation packages often address parts of the problem successfully, there remain certain shortcomings such as the inability to incorporate non-continuous input data and nonlinear transaction costs.

A core reason for the "hardness" of the portfolio problem is the sheer number of possible portfolios, making solution by enumeration a daunting task. The horrors of enumeration can be illustrated as follows.

Say we have a universe of N assets from which to form an optimal portfolio consisting of a smaller number of assets, say K. The number of possible combinations is

$$C_K^N = \binom{N}{K} = \frac{N!}{K!(N-K)!}$$

Now for each K-asset portfolio assume the asset weights are defined with a resolution of $r$, so for example if $r = 1$ the asset's weighting is 100% (or 0%), if $r = 2$ its weighting is either 50% or 100% (or 0%). (The number of weighting possibilities is given by $r+1$ and the percentage resolution is given by $p = \frac{100}{r}$, so a weighting with a percentage resolution of 1% will require $r = 100$). Clearly, the total number of possible portfolios with different combinations of asset weights is given by $K^{r+1}$.

However, only a subset of these combinations will have asset weights that sum to 100%. This is known as C'(n,k), a *k-composition of n*, which is a partition of n into exactly k parts, *with regard to order,* where each part is an integer greater than or equal to zero. The number of compositions is given by C'(n,k) = $C_{k-1}^{n+k-1}$.

The total number of enumeration possibilities E is therefore given by

$$E = C_K^N \cdot C'(r+K\text{-}1, K\text{-}1) = C_K^N \cdot C_{K-1}^{r+K-1}$$

$$= \frac{N!}{K!(N-K)!} \cdot \frac{(r+K-1)!}{(K-1)!(r)!}$$

We will be searching for the optimal 40-stock portfolio selected from a universe of 100 shares, and wish weightings to be defined within 1%. Therefore $N = 100$, $K = 40$ and $p = 1$, giving $r = 100$.

So $E = C_{40}^{100} \cdot C_{39}^{139}$

$$= \left(\frac{100!}{40!60!}\right) \cdot \left(\frac{139!}{39!100!}\right) = (1,4 \times 10^{28}) \cdot (5,1 \times 10^{34}) = 6,9 \times 10^{62} \text{ portfolios}$$

The latest Cray T3E supercomputer operates at 2,4 teraflops. Assume that the evaluation of each portfolio will require around 300 floating-point operations. Therefore to evaluate each portfolio the Cray will take $\frac{300 \text{ flop/portfolio}}{2,4 \times 10^{12} \text{ flop/second}} =$ $1,25 \times 10^{-10}$ seconds/portfolio (or will process $8 \times 10^9$ portfolios/second). The time required to evaluate all the possible portfolios is therefore $(1,3 \times 10^{-10} \text{ sec/portfolio}) \cdot (6,9 \times 10^{62} \text{ portfolios}) = 8,7 \times 10^{52}$ seconds or $2,7 \times 10^{45}$ years.

The latest estimates for the age of the universe are only around $1,3 \times 10^{10}$ years.

Optimisation by enumeration could be tedious.

## 1.2   Objectives

The objective of the research is to investigate the ability of metaheuristic methods to deliver high-quality solutions for the mean-variance model when enriched by additional practical constraints. We therefore develop a model which reflects the most important real-life aspects of portfolio optimisation and investigate its solution by two heuristic methods: genetic algorithms (GA) and tabu search (TS).

The Markowitz model is extended by the incorporation of:
♦   floor and ceiling constraints;
♦   nonlinear transaction costs; as well as
♦   cardinality constraints.

With powerful and cheap computation now widely available, heuristic approaches are attractive, as they are independent of the objective function and the structure of the model and its constraints, while also being general and robust.

## 1.3 Problem description

In the original Markowitz model it was assumed that asset class returns follow a multivariate normal distribution. The return on a portfolio of assets therefore can be described completely by the first two moments, i.e. the expected or mean return and the variance of these returns (the measure of risk). Optimisation consists of finding the set of portfolios which provide the lowest level of risk for any required level of return or, conversely, the highest return for any specified level of risk. This set of portfolios is called the efficient frontier and may be found exactly by quadratic programming (QP). It is usually displayed as a curve plotting the expected portfolio returns against the standard deviation of each of these forecast returns.

There are essentially two justifications for the mean-variance assumption. Either preferences are quadratic in consumption or asset prices are jointly normally distributed [28]. A weakness of the model is this assumption of multivariate normality. Distributions of asset returns have been shown to be leptokurtotic, i.e. with a higher probability of extreme values (e.g. Mills [2]). Theoretically this means that the first two moments, of expected return and variance, are insufficient to describe the portfolio fully and higher moments are required. The model also states that each investor can assign a welfare, or utility, score to competing investment portfolios based on the expected return and risk of those portfolios. There is thus the assumption that these first two moments, of expected return and risk, are sufficient to determine an investor's utility function, usually represented by an indifference curve. If asset class returns are not normally distributed, investor utility could be represented by very different distributions which nevertheless have the same mean and standard deviation. A useful extension of the model would therefore be to allow the investor to choose between these two distributions.

There are also "floor" and "ceiling" constraints in practical portfolio construction. Extremely small weightings of an asset will have no effective influence on the portfolio's return but will add to administrative and monitoring costs, so floor or minimum weightings are commonly established. Similarly, very high weightings in any one asset introduce excessive exposure to the idiosyncrasies of that asset (even though the portfolio's overall risk may appear acceptable) and a policy "ceiling" on assets or asset classes is often set. In addition, in certain types of portfolio further legal and regulatory limits on asset class weightings exist. For example, unit trusts generally are required to have a minimum of 5% in cash, not more than 75% in equities and not more than 20% in offshore assets. Again, incorporation of these constraints in the Markowitz model is difficult.

The simplest situation exists when the nonnegativity constraints on the asset class weights are omitted from the basic model (thus allowing short sales). In this case, a closed-form solution is easily obtained by classical Lagrangian methods and various approaches have been proposed to increase the speed of resolution for the computation of the whole mean-variance frontier or the computation of a specific portfolio combined with an investment at the risk-free interest rate. The problem becomes more complex when the nonnegativity constraints are added to the formulation. The addition of these nonnegative weightings and any floor and ceiling constraints results in a QP problem which can still be solved efficiently by specialised algorithms such as Wolfe's adaptation of the simplex method [3]. However, as the number of assets increases the problem becomes increasingly hard to manage and solve and ad hoc methods are required to take advantage of the sparsity or of the special structure of the covariance matrix, as discussed by Perold [4].

It has been shown by the capital asset pricing model (CAPM) (see e.g. Sharpe [5]) and arbitrage pricing theory (APT) that the systematic risk of a portfolio, i.e. the portion of risk dependent only on the market, is bounded from above by the average of the portfolio assets' specific variances divided by the number of assets in the portfolio; it therefore declines rapidly and asymptotically to this limit as the number of stocks increases. Empirically, in practice systematic risk becomes negligible when the number of assets in the portfolio exceeds approximately 20-25 securities. (There is, however, evidence [29] that in recent years this number may have increased substantially, to around 50 stocks.) In addition, the costs of following a large number of assets is substantial, so the number of assets in a portfolio is usually limited to a very small subset of the available universe, normally in the region of 30-50 stocks in an equity portfolio (compared with a universe of around 600 stocks currently listed on the JSE). This type of cardinality constraint is not easily applied to the Markowitz model as it results in a mixed integer nonlinear programming problem, and classical algorithms are typically unable to optimise this problem.

The issue of transaction costs is critical to the construction and management of portfolios and the impact of these costs on performance can be major. Costs are, firstly, not fixed and in addition to a fixed charge usually comprise a proportional element as well as various taxes. Secondly, there is an additional "liquidity premium" which must be paid in the case of large orders in stocks that suffer from limited tradability. This estimated liquidity premium is strongly nonlinear and can be up to two orders of magnitude larger than the negotiated costs. Transaction costs are therefore often of major concern to large institutional investors. The precise treatment of transaction costs leads to a nonconvex minimisation problem for which there is no efficient method of calculating an exact optimal solution. In addition, most approaches to the incorporation of costs in the mean-variance model ignore the nonlinearity and therefore add little value.

## 1.4 Literature review and previous work

Patel [6] showed that even for fixed transaction costs their exclusion from a portfolio selection model often leads to an inefficient portfolio in practice. Although Perold [4] and Mulvey [7] approximate a transaction cost function by a piecewise linear convex function, this is not valid for the nonconvex shape we estimate for the actual transaction cost function.

More recently, Konno and Yamazaki [8] proposed a linear programming model using the mean-absolute deviation (MAD) as the risk function. The model assumes no particular distribution for asset returns and is equivalent to the Markowitz model when they have a multivariate normal distribution. This model has been applied where there are asymmetric return distributions, such as in a mortgage-backed securities portfolio optimisation (Zenios and Kang [9]). The possible asymmetry of returns is taken into account by Konno, Shirakawa and Yamazaki [10], who extended the MAD approach to include skewness in the objective function. Konno and Suzuki [11] considered a mean-variance objective function extended to include skewness. Finally, Konno and Wijayanayake [12] use a branch-and-bound algorithm to solve the MAD optimisation model for a concave cost function which is approximated by linear segments. However, minimum lot sizes were not incorporated, even though the cost curve is concave in the area of small transactions. For large transactions the cost curve is believed to be convex, and this is the area of interest to large institutional investors.

Xia, Liu, Wang and Lai [13] addressed the situation where the order of expected returns is known and solved this new portfolio optimisation model with a genetic algorithm. The effect of transaction costs was also examined, but only for the case of proportional costs. Loraschi, Tettamanzi, Tomassini and Verda [14] presented a distributed genetic algorithm for the unconstrained portfolio optimisation problem based on an island model where a genetic algorithm is used with multiple independent subpopulations, while Crama and Schyns [15] developed a model incorporating floor and ceiling constraints, turnover constraints, trading constraints (i.e. minimum trading lot sizes) and cardinality constraints which was solved by a simulated annealing algorithm. Costs, however, were ignored. The algorithm is versatile, not requiring any modification for other risk measures while the algorithms of Perold [4] and Bienstock [16] explicitly exploit the fact that the objective function is quadratic and that the covariance matrix is of low rank.

The mixed-integer nonlinear (quadratic) programming problem which arises from the incorporation of cardinality constraints can be solved by adapting existing algorithms. For example, Bienstock [16] uses a branch-and-bound algorithm while Borchers and Mitchell [17] use an interior point nonlinear method.

Alternatively, the quadratic risk function in the Markowitz model can be approximated by a linear function, enabling mixed-integer linear programming to be used. Speranza [18] showed that taking a linear combination of the mean semi-absolute deviations (i.e. mean deviations below and above the portfolio rate of return) resulted in a model equivalent to the MAD model. In Speranza [19] this linear model was extended to incorporate fixed and proportional transaction costs as well as cardinality and floor constraints. Despite the model's underlying linearity a heuristic algorithm had to be tailored for its solution, and it was not possible to solve the model in reasonable time if the number of stocks was greater than 15-20. In practical problems, more general and robust heuristic methods would be an advantage. Manzini and Speranza [20] used the same approximation to consider floor constraints or minimum lots. While minimum lots may be relevant to small individual investors they are of little interest to large institutional investors.

Chang, Meade, Beasley and Sharaiha [21] constructed a cardinality-constrained Markowitz model incorporating floor and ceiling constraints which was solved using genetic algorithms, simulated annealing and tabu search, but costs were not addressed.

Tabu search (TS) was developed by Glover [22] and was applied by Glover, Mulvey and Hoyland [23] to a portfolio optimisation problem involving dynamic rebalancing to maintain constant asset proportions, using a scenario approach to model forecast asset returns.

While many of these approaches have combined various real-life constraints in their model formulations, in none of the above previous work have floor and ceiling constraints, nonlinear transaction costs and cardinality constraints all been incorporated simultaneously in one model.

The exploration and solution of the optimisation problem will follow these steps:

1. Determine a realistic cost function.

2. Using a relatively small ten-asset portfolio, establish the impact on portfolios of:
   - floor and ceiling constraints
   - costs

- combined floor and ceiling constraints and costs

These portfolios can be solved by traditional nonlinear solvers such as LINGO or Excel's Solver.

3. Establish the credentials of the heuristic method by now finding the efficient frontier for a large 100-stock portfolio with floor and ceiling constraints and costs using both traditional solvers and genetic algorithms (GAs) and comparing the results. If GAs provide acceptable results, proceed.

4. Add the cardinality constraint to the model. It is now no longer solvable by the traditional methods.

5. Solve this complete, large model using GAs. Construct the efficient frontier.

6. Estimate the risk-aversion parameter $w$ from this efficient frontier.

7. Use this value of $w$ to optimise actual portfolios.

# 2. Theory and problem formulation

## 2.1 Unconstrained Markowitz model

If

$N$ = the number of assets in the investable universe

$R_i$ = the expected return of asset $i$ ($i = 1; ...; N$) above the risk-free rate $r_f$

$\sigma_{ij}$ = the covariance between assets $i$ and $j$ ($i = 1; ...; N$, $j = 1; ...; N$)

$R_p$ = the expected return of the portfolio above the risk-free rate

$x_i$ = the weight in the portfolio of asset $i$ ($i = 1; ...; N$), where $0 \leq x_i \leq 1$

then the portfolio's expected return is given by

$$R_p = \sum_{i=1}^{N} R_i x_i \qquad - (1)$$

and its risk is given by the variance of expected returns

$$\sigma_p^2 = \sum_{i=1}^{N} \sum_{j=1}^{N} \sigma_{ij} x_i x_j \qquad - (2)$$

The unconstrained portfolio optimisation problem is therefore

$$\text{minimise} \sum_{i=1}^{N} \sum_{j=1}^{N} \sigma_{ij} x_i x_j$$

or, using the fact that $\sigma_{ij} = \rho_{ij} s_i s_j$, where $\rho_{ij}$ is the correlation between $i$ and $j$ and $s_i$, $s_j$ represent the standard deviation of their returns (usually monthly, annualised),

$$\text{minimise} \sum_{i=1}^{N} \sum_{j=1}^{N} \rho_{ij} s_i s_j\, x_i x_j \qquad - (3)$$

$$\text{subject to } R_p = \sum_{i=1}^{N} R_i x_i$$

$$\sum_{i=1}^{N} x_i = 1 \qquad - (4)$$

$$0 \le x_i \le 1 \quad i = 1, \ldots N$$

The portfolio's variance or risk is therefore minimised for a required rate of return $R_p$, while all asset weights sum to one. Note that in Markowitz' original article returns referred to returns in excess of the risk-free rate (which is often overlooked by practitioners), and this definition is used in the model. This is a simple nonlinear (quadratic) programming problem which is easily solved using standard techniques.

In this form the model requires $(n^2 + 3n)/2$ items of data for an $n$-asset portfolio, comprising $n$ estimates of expected returns, $n$ estimates of variances and $(n^2-n)/2$ estimates of correlations (since the correlation matrix's diagonal elements are all one and $\rho_{ij} = \rho_{ji}$). Therefore a portfolio consisting of only 50 assets requires 1325 separate items of data while a 100-asset portfolio would require 5150 data items.

However, assuming that the only reason for the assets' correlation is their common response to market changes, and that the assets are stocks, the measure of their correlation can be obtained by relating the returns of the stock to the returns of a stock market index, usually that of the overall market, as shown by Sharpe [5]. The returns of a stock can then be broken into two components, with one part resulting from the market and the other independent of the market, as follows:

$$R_i = \alpha_i + \beta_i R_m + e_i \qquad - (5)$$

where $\alpha_i$ is the component of security $i$'s return which is independent of the market's performance, $R_m$ is the return of the market index, $\beta_i$ is a constant that measures the expected change in $R_i$ for a given change in $R_m$ and $e_i$ is a random error or "firm-specific" component. Note again that returns $R_i$ and $R_m$ are returns in excess of the risk-free rate $r_f$. For each asset $i$, on a graph of $R_i$ versus $R_m$, $\beta_i$ is the slope of the regression line, $\alpha_i$ is the intercept and the residuals $e_i$ are the deviations from the regression line to each point. This is implicitly a disequilibrium model, since a market in equilibrium would require no excess returns, or $\alpha_i = 0$.

Therefore the expected return of the portfolio and the variance of this expected return can be simplified as

$$R_p = \alpha_p + \beta_p R_m \qquad - (6)$$

and

$$\sigma_p^2 = \beta_p^2 \sigma_m^2 + \sum_{i=1}^{N} x_i^2 \sigma_{ei}^2 \qquad - (7)$$

where $\sigma_{ei}^2$ is the variance of the random error component $e_i$, $\sigma_m^2$ is the variance of $R_m$ and $\alpha_p$ and $\beta_p$ are given by

$$\alpha_p = \sum_{i=1}^{N} x_i \alpha_i \qquad \text{- (8)}$$

$$\beta_p = \sum_{i=1}^{N} x_i \beta_i. \qquad \text{- (9)}$$

Therefore $\sigma_p^2 = \left[\sum_{i=1}^{N} x_i \beta_i\right]^2 \sigma_m^2 + \sum_{i=1}^{N} x_i^2 \sigma_{ei}^2 \qquad \text{- (10)}$

For each asset class $i$ the variance of residuals $\sigma_{ei}^2$ is found as follows. If the residual of each point for asset class $i$ is $e_{it,}$ because the mean of $e_{it}$ is zero, $e_{it}^2$ is the squared deviation from its mean. The average value of $e_{it}^2$ is therefore the estimate of the variance of the firm-specific component. Dividing the sum of squared residuals by the degrees of freedom of the regression (which for $T$ points is $T$-2) gives an unbiased estimate of $\sigma_{ei}^2$.

So $\sigma_{ei}^2 = \dfrac{\sum_{t=1}^{T} e_{it}^2}{T-2} \qquad \text{- (11)}$

It may be noted that for a large number of stocks in the portfolio, usually when $i > 25$, which is the situation which will be analysed in this research, the firm-specific variances will tend to cancel out and their sum will tend towards zero. This is because these $e_i$ are independent and all have zero expected value, so as more stocks are added to the portfolio the firm-specific components tend to cancel out, resulting in ever-smaller non-market risk. The portfolio's variance will therefore comprise only so-called systematic risk $\left[\sum_{i=1}^{N} x_i \beta_i\right]^2 \sigma_m^2$ in the above equation for $\sigma_p^2$.

This single-index model reduces the estimated input data from $(n^2 + 3n)/2$ to $3n + 1$ data items, comprising $n$ expected returns $R_i$, $n$ forecast betas $\beta_i$, $n$ estimates of the firm-specific variances $\sigma_{ei}$ and one estimate of the market's variance $\sigma_m$. The data

12$$\alpha_p = \sum_{i=1}^{N} x_i \alpha_i \qquad \text{- (8)}$$

$$\beta_p = \sum_{i=1}^{N} x_i \beta_i. \qquad \text{- (9)}$$

Therefore $\sigma_p^2 = \left[\sum_{i=1}^{N} x_i \beta_i\right]^2 \sigma_m^2 + \sum_{i=1}^{N} x_i^2 \sigma_{ei}^2 \qquad \text{- (10)}$

For each asset class $i$ the variance of residuals $\sigma_{ei}^2$ is found as follows. If the residual of each point for asset class $i$ is $e_{it,}$ because the mean of $e_{it}$ is zero, $e_{it}^2$ is the squared deviation from its mean. The average value of $e_{it}^2$ is therefore the estimate of the variance of the firm-specific component. Dividing the sum of squared residuals by the degrees of freedom of the regression (which for $T$ points is $T$-2) gives an unbiased estimate of $\sigma_{ei}^2$.

So $\sigma_{ei}^2 = \dfrac{\sum_{t=1}^{T} e_{it}^2}{T-2} \qquad \text{- (11)}$

It may be noted that for a large number of stocks in the portfolio, usually when $i > 25$, which is the situation which will be analysed in this research, the firm-specific variances will tend to cancel out and their sum will tend towards zero. This is because these $e_i$ are independent and all have zero expected value, so as more stocks are added to the portfolio the firm-specific components tend to cancel out, resulting in ever-smaller non-market risk. The portfolio's variance will therefore comprise only so-called systematic risk $\left[\sum_{i=1}^{N} x_i \beta_i\right]^2 \sigma_m^2$ in the above equation for $\sigma_p^2$.

This single-index model reduces the estimated input data from $(n^2 + 3n)/2$ to $3n + 1$ data items, comprising $n$ expected returns $R_i$, $n$ forecast betas $\beta_i$, $n$ estimates of the firm-specific variances $\sigma_{ei}$ and one estimate of the market's variance $\sigma_m$. The data requirements for a 50-asset and a 100-asset portfolio are reduced dramatically from the previous case, to only 151 and 301 data items respectively.

Equations (3) and (4) can equivalently be solved by maximising portfolio return $R_p$ for a required level of risk $\sigma_p^2$. Normally one is trying to optimise a combination of both returns and risk, and it is standard practice (see e.g. [21]) to introduce a weighting parameter $w$ to form a new objective function which is a weighted combination of both return and risk. This so-called risk-aversion parameter $w$ ($0 \leq w \leq 1$) enables the efficient frontier to be traced out parametrically. The problem can therefore be restated as

$$\text{maximise} \quad (1-w) \sum_{i=1}^{N} R_i x_i - w \sum_{i=1}^{N} \sum_{j=1}^{N} \sigma_{ij} x_i x_j \quad - (12)$$

or, introducing the new formulation,

$$\text{maximise} \quad (1-w) \sum_{i=1}^{N} R_i x_i - w \left[ \left\{ \sum_{i=1}^{N} x_i \beta i \right\}^2 \sigma_m^2 + \sum_{i=1}^{N} x_i^2 \sigma_{ei}^2 \right] \quad - (13)$$

Solving the last (QP) equation (13) for various values of $w$ results in combinations of portfolio return and variance which trace out the efficient frontier. Finding these points on the efficient frontier which represents optimal combinations of return and risk is exactly the same as solving equations (1), (3) and (4) for varying values of $R_p$. This curve represents the set of Pareto-optimal or non-dominated portfolios.

When $w = 0$, returns are paramount and risk is not taken into consideration. The portfolio will consist of only a single asset, the one with the highest return. The condition $w = 1$ represents the situation where risk is minimised irrespective of return. This will usually result in portfolio consisting of many assets, since it is the combination of assets and the lack of correlation between them that reduces the portfolio's risk to below the level of any individual asset. Most investors' risk preference will lie somewhere between these two extremes.

## 2.2 Constraints

The real-world extensions to the model can now be introduced.

### 2.2.1 Floor and ceiling constraints

In practical portfolio optimisation both floor and ceiling constraints need to be addressed. Floor constraints are implemented in practice to avoid excessive administration costs for very small holdings which will have a negligible influence on the portfolio's performance, while ceiling constraints are set on the principle that excessive exposure to any one portfolio constituent needs to be limited as a matter of policy.

If  $a_i$ = the minimum weighting that can be held of asset $i$ ($i = 1, \ldots N$)
    $b_i$ = the maximum weighting that can be held of asset $i$ ($i = 1, \ldots N$)
then the constraint is simply formulated as

$$a_i \leq x_i \leq b_i \qquad - (14)$$

where $0 \leq a_i \leq b_i \leq 1$ ($i = 1, \ldots N$)

It may be noted that the floor constraints generalise the nonnegativity constraints imposed in the original model. Various researchers have incorporated floor and/or ceiling constraints in their models ([15], [19], [20], [21]).

### 2.2.2 The cost function

No attempts to model transaction costs comprehensively were found in the literature. Costs can be large in comparison with portfolio returns, particularly in sideways-moving and illiquid markets, and realistic modelling can be critically important. The problem has been addressed as follows.

The conceptual shape of the transaction cost function is shown in Figure 1, where units can refer to either number of shares or deal size in monetary units.

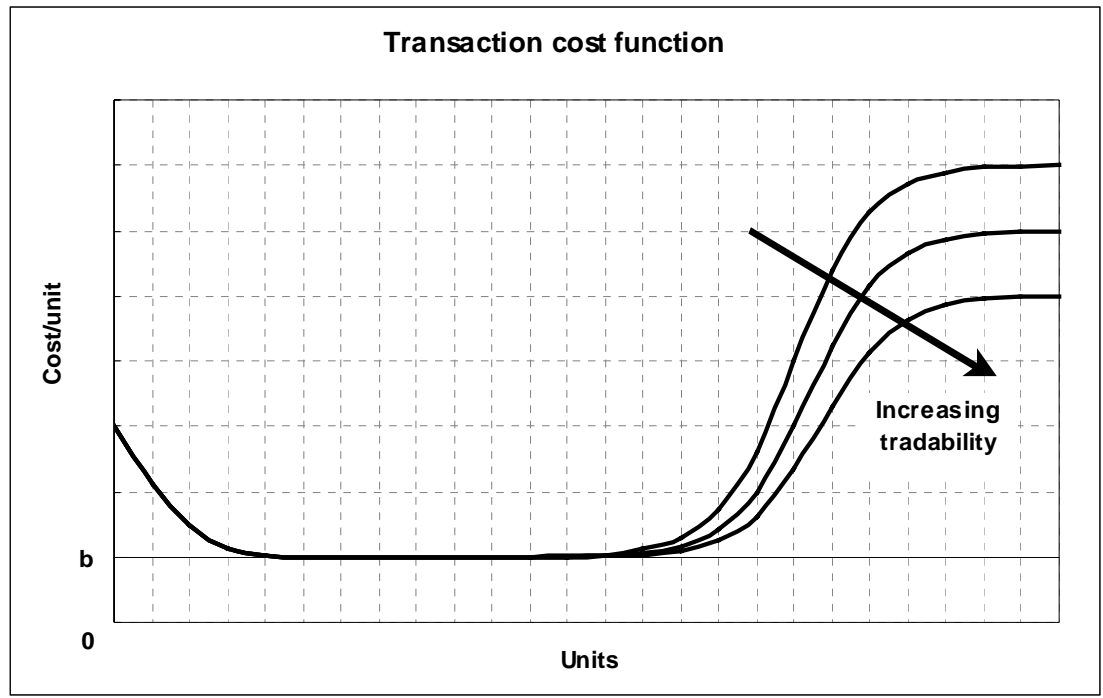

*Figure 1: Illustrative transaction cost functions*

Ignoring other fixed costs and taxes, for most deal sizes the unit cost equals the brokerage rate $b$. However, deals that are smaller than "round lots" of 100 shares attract an additional cost, the "small size" premium, while large deals also attract an additional cost, the "illiquidity premium" or "impact cost". This discussion will be restricted to the high end of the cost curve, as this is the region relevant to institutional investors.

If

$m$ = marketable securities tax (MST) rate

$f$ = fixed charge component

$v$ = value-added tax (VAT) rate

$b$ = brokerage rate

$s$ = transaction value

$t$ = asset tradability (average value traded per time period)

$p$ = illiquidity premium

$C$ = total transaction cost

$c'$ = total unit transaction cost

then the total transaction cost is given by

$$C = (1 + v)[f + (b + p)s] + ms$$

$$= (1 + v)f + [(1 + v)(b + p) + m]s \qquad - (15)$$

and the total unit transaction cost is

$$c' = C/s = (1 + v)f/s + [(1 + v)(b + p) + m] \qquad - (16)$$

Note that the illiquidity premium can be introduced into equation (15) in any form; for convenience we have elected to consider it an increment to the brokerage rate.

Interviews with market dealers established that the illiquidity premium is overwhelmingly a function of deal size relative to the shares' tradability and the period over which the deal is done. Clearly, spreading a deal over time will reduce the market impact cost premium. The cost function will be estimated on the basis that deals are not spread over time. This will represent the upper limit of the illiquidity premium and any spreading will therefore tend to reduce the calculated costs.

The illiquidity premium is therefore given by a function $F$ of $s/t$,

$$p = F(s/t).$$

The influence of other factors is relatively negligible. Using the dealers' estimates for the size of the illiquidity premium for various values of $(s/t)$ shows that it initially rises rapidly against this variable but then slows asymptotically to an upper limit as shown in Figure 1.

Note that it has been assumed that the illiquidity premium function is smooth. This may not necessarily be true in a "lumpy" market, where a small increase in proffered deal size could trigger the release of a large quantity of stock from a specific seller. The illiquidity premium function for each portfolio constituent may therefore not be smooth. However, it has been assumed that for the portfolio in aggregate this function is indeed smooth. It may be noted that even if such a discontinuous function could be determined, which is highly unlikely, the advantage of metaheuristic methods is that the optimal portfolio can still be found.

This ramp function can be modelled by any of the following functions:

Hyperbolic tangent:    $p(x) = a \tanh c(x-d)$    (3 parameters)

Logistic equation:    $p(x) = \dfrac{a}{1+ke^{-c(x-d)}}$    (4 parameters)

Single-term exponential:    $p(x) = a[1-ke^{-c(x-d)}]$    (4 parameters)

Two-term exponential:    $p(x) = a[1-ke^{-g(x-d)} - be^{-c(x-d)}]$    (6 parameters)

where $x = s/t$ and $d$ is a lag parameter.

While all of these functions have the required shape, the logistic equation does not meet the requirement that $p(x) = 0$ at $x = 0$ (ignoring the "small size" premium), and of the remaining three equations only the two-term exponential function has the additional property that $\dfrac{dp}{dx} = 0$ at $x = 0$, which correctly reflects the situation shown in Figure 1, again disregarding the "small size" premium.

Applying the condition $p(0) = 0$ to this equation leads to $b = 1-k$, while the second condition $\left.\dfrac{dp}{dx}\right|_{x=0} = 0$ results in the requirement that $g = (k-1)/k$. It was also found that for practical purposes $d=0$ as shifting the curve generally does not produce a better fit to the empirical data. The number of parameters in this equation is therefore reduced from six to three.

A two-term exponential function of the form

$$p(s/t) = a[1 - ke^{-((k-1)/k)\,c(s/t)} + (k-1)e^{-c(s/t)}] \qquad \text{- (17)}$$

is therefore used.

This curve is fitted empirically to the market dealers' estimates of $p$ for various values of $s/t$ using the three parameters $a$, $k$ and $c$. These parameters are selected to give the best fit by minimising the sum-of-squares error of the fitted curve. An example of a fitted cost curve is shown in Table 2 and Figure 3, both on page 22.

### 2.2.3 Cardinality constraint

Cardinality constraints are combinatorial in nature. Some researchers, e.g. Chang et al [21], have incorporated cardinality constraints in their models.

Define the cardinality variable as $z_i$,

where $z_i = 1$    if any amount of asset $i$ ($i = 1, \ldots N$) is held      - (18)
        $= 0$   otherwise

     $K =$ the maximum number of assets allowed in the portfolio

Then the cardinality constraint becomes

$$\sum_{i=1}^{N} z_i = K \qquad \text{- (19)}$$

where $K \leq N$

and    $z_i \in [0,1]$ is the integrality constraint.

Note that if any specific asset is required to be in the portfolio, this is achieved simply by setting $z_i = 1$ for that asset prior to optimisation.

The cardinality constraint now needs to be combined with the floor and ceiling constraint, since $a_i$ and $b_i$ are now the minimum and maximum weightings that can be held of asset *i* <u>only if</u> any of asset *i* is held. The floor and ceiling constraint (14) therefore becomes

$$a_i z_i \leq x_i \leq b_i z_i$$

with the additional condition that

$$1/b_i \leq K \leq 1/a_i \qquad - (20)$$

In our example $b_i=b$ and $a_i=a$ for all *i*; individual limits could, however, be set for each asset class.

This cardinality constraint may also be set to a range, i.e.

$$K_l \leq \sum_{i=1}^{N} z_i \leq K_u \qquad - (21)$$

where $K_l$, $K_u$ represent the lower and upper limits on the number of assets in the portfolio respectively.

# 3. Solution methods

## 3.1 Problem definition

**Efficient frontier**

The construction of an efficient frontier is illustrated for a two-asset portfolio in Table 1 and Figure 2, both on page 20.

| Two-asset portfolio | | | | | | |
|---|---|---|---|---|---|---|
| | | | $r_f =$ | 13.0 | $\sigma_m =$ | 0.276 |
| Asset no. $i$ | Asset name | Weight $x_i$ (frac) | Forecast return $r_i$ (%) | Excess return $R_i$ (%) | Beta $\beta_i$ (x) | Forecast $\sigma_{ei}$ (%) |
| 1 | A | 0.50 | 50.0 | 37.0 | 1.10 | 0.150 |
| 2 | B | 0.50 | 10.0 | -3.0 | 0.90 | 0.200 |
| Portfolio sum/average | | 1.00 | 30.0 | **17.0** | 1.00 | **0.303** |

*Table 1: Two-asset portfolio data*

Forecast asset returns of 50% and 10% for assets A and B become excess returns of 37% and -3% above the risk-free rate of 13% respectively. The portfolio return is a linear combination of the asset class returns, as given by equation (1). However, the portfolio's risk level is not a linear combination of the asset class risks due to the nonlinear term in equation (10). For example, the 50:50 combination of asset classes shown in Table 1 results in a risk level below that of either of the individual assets.

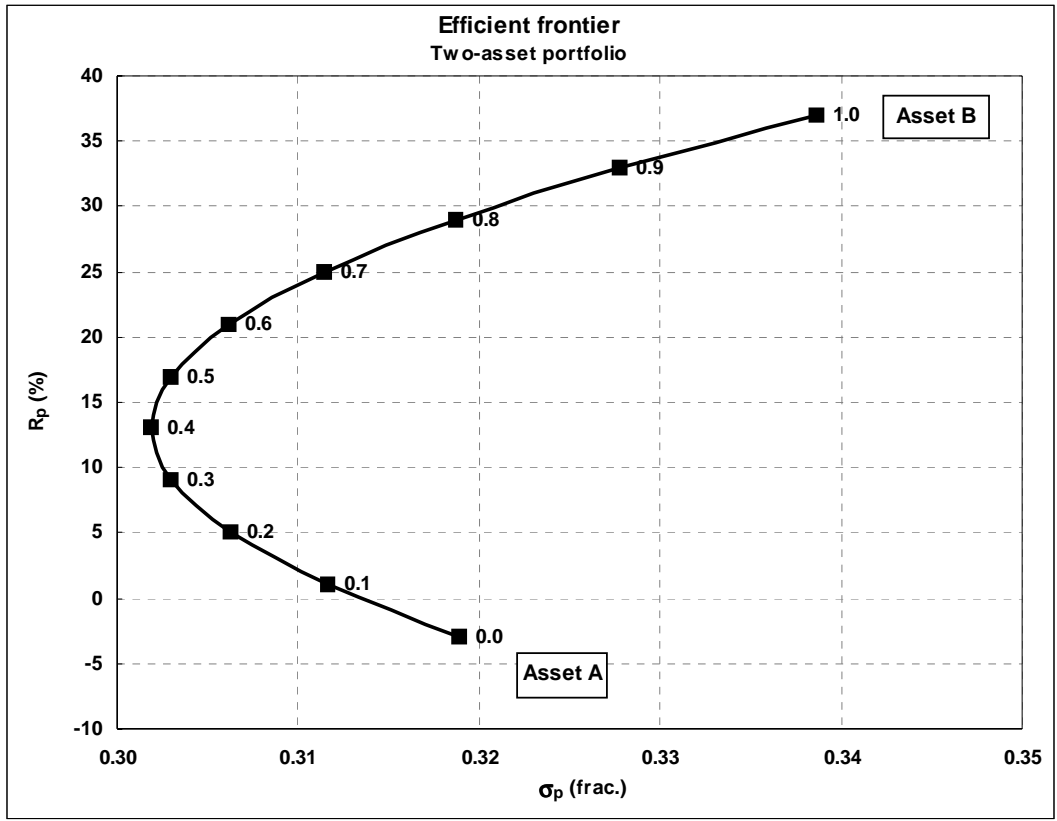

*Figure 2: Two-asset efficient frontier*

Only the upper portion of the curve in Figure 2 will be considered in the following sections. It is by definition the efficient frontier, since the bottom half represents lower returns than on the upper half for any given level of risk.

For the complete portfolio model the upper portion of the efficient frontier is calculated in all cases by varying $w$ in the objective function represented by the expression (13). In this objective function the variables $R_i$, $\beta_i$, $\sigma_m$ and $\sigma_{ei}$ are all known, so it can be maximised by finding the optimal combination of assets $x_i$ ($i = 1, \ldots N$). The return and risk of this portfolio are represented by the components of the objective function defined by equations (1) and (10) and determine the point on the efficient frontier associated with that value of $w$.

**Cost function**

The illiquidity premium (equation 17) was modelled as follows.

The illiquidity premium for various deal sizes was estimated by interviewing both market dealers and selected institutional fund managers. The interviewing technique was direct questioning and an unweighted average of the responses was used.

The average estimated values of $p$ for various values of ($s/t$) as determined by these interviews are shown in Table 2 and Figure 3, both on page 22. A value of $t =$ R300m/month has been used throughout; it can be individualised for each asset if required.

Empirically fitting the parameters to minimise the least squares errors results in the parameter values $a$, $k$ and $c$ for equation (17) which are shown in Table 2, and the resulting curve for the illiquidity premium is also shown in Figure 3.

| Fitted cost function | | | | | | |
|---|---|---|---|---|---|---|
| Parameters: | | a | k | c | | |
| | | 30.45 | 100.0 | 1.246 | | |
| s | s/t | Estimated p | Fitted p | b+p | Total cost C | Unit cost c' |
| (Rm) | (months) | (%) | (%) | (%) | (Rm) | (%) |
| 0 | 0.0 | 0 | 0.0 | 0.0 | 0.0 | - |
| 150 | 0.5 | 5 | 3.9 | 4.2 | 7.6 | 5.0 |
| 300 | 1.0 | 11 | 10.7 | 11.0 | 38.4 | 12.8 |
| 450 | 1.5 | 14 | 16.9 | 17.2 | 89.3 | 19.8 |
| 600 | 2.0 | 22 | 21.6 | 21.9 | 151.1 | 25.2 |
| 750 | 2.5 | 24 | 24.8 | 25.1 | 216.6 | 28.9 |
| 900 | 3.0 | 30 | 27.0 | 27.3 | 281.9 | 31.3 |
| 1050 | 3.5 | 30 | 28.3 | 28.6 | 345.2 | 32.9 |
| 1200 | 4.0 | 30 | 29.2 | 29.5 | 406.2 | 33.8 |
| 1350 | 4.5 | 30 | 29.7 | 30.0 | 464.9 | 34.4 |
| 1500 | 5.0 | 30 | 30.0 | 30.3 | 521.9 | 34.8 |

*Table 2: Fitted illiquidity premium function*

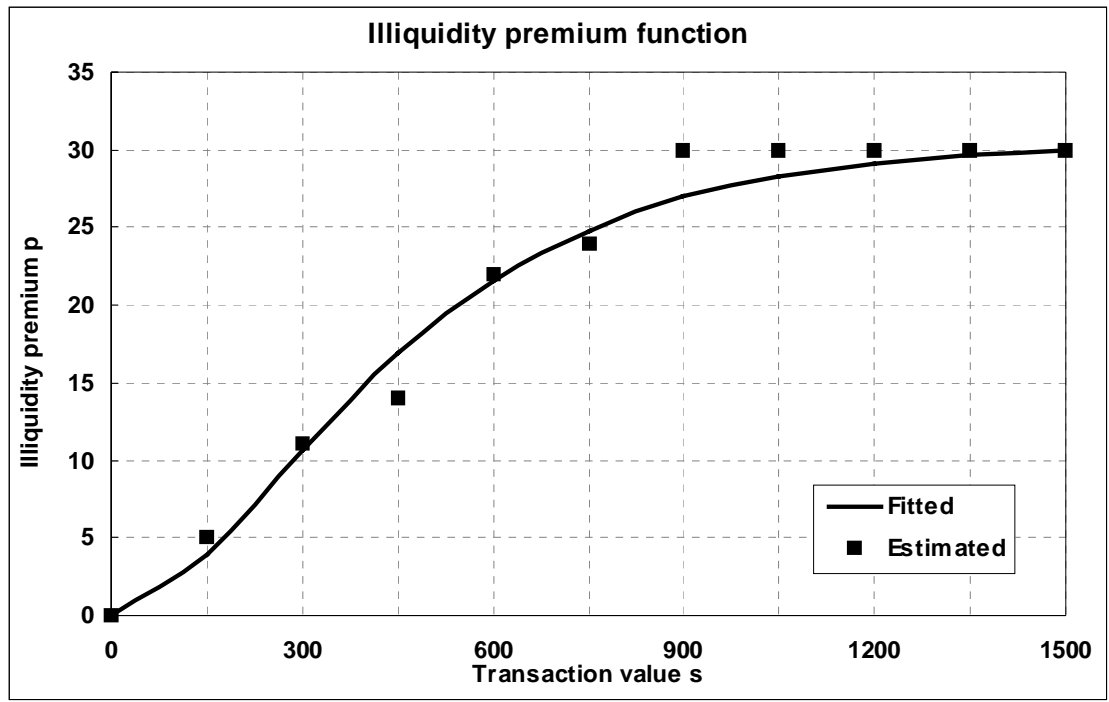

*Figure 3: Fitted illiquidity premium*

The set of curves for various values of *t* is shown in Figure 4 on page 23.

**Illiqidity premium surface**

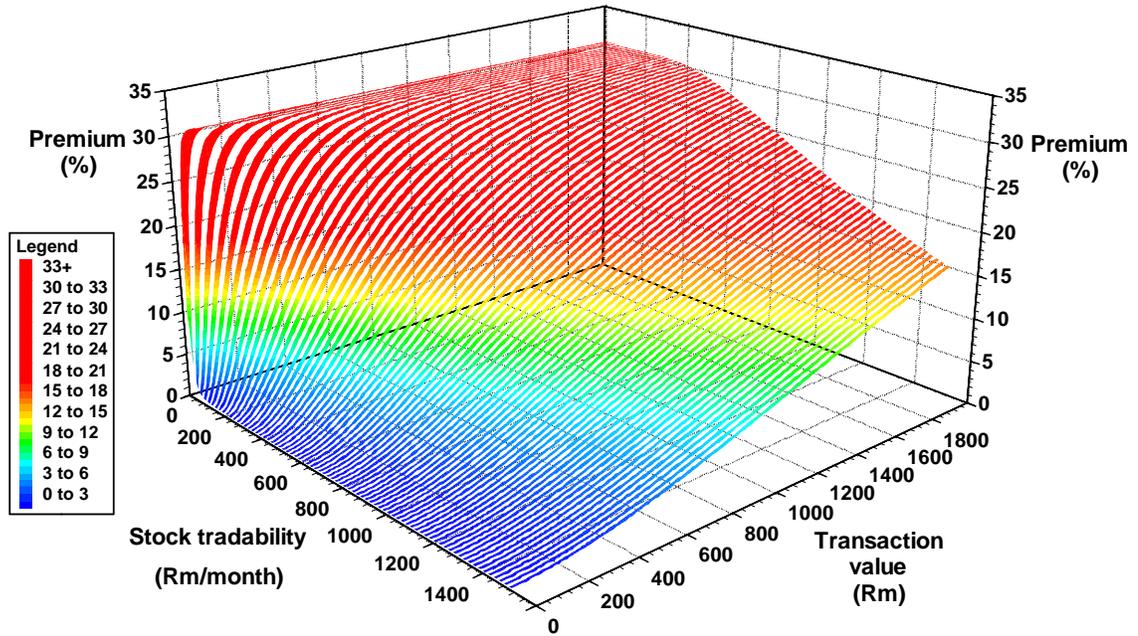

*Figure 4: Illiqidity premium surface*

As asset tradability increases the cost curve both declines and becomes more linear.

The values used for the remaining variables in the total unit cost equation (15) are shown in Table 3.

| Cost elements | | | | |
|---|---|---|---|---|
| | Variable | | Units | Value |
| m | = | MST rate | % | 0.25 |
| f | = | fixed minimum charge | R | 15 |
| v | = | VAT rate | % | 14 |
| b | = | brokerage rate | % | 0.3 |
| t | = | asset tradability | Rm/month | 300 |

*Table 3: Transaction cost parameters*

The total unit cost as given by equation (16) therefore becomes

$c' = C/s = (1+14/100)(15)/s + [(1+14/100)(0{,}30+ p) + 0{,}25]$

where

$$p = 30{,}45 \,[1 - 100 e^{-((0{,}990)(1{,}246)/300)s} + 99 e^{-(1{,}246/300)s}]$$

Therefore

$$c' = 17{,}1/s + 35{,}30 - 3044{,}6 e^{-0{,}004112 s} + 3014{,}2 e^{-0{,}004153 s} \qquad - (22)$$

This total unit cost function is shown in Figure 5. Note costs approach the upper limit of 35,3 asymptotically.

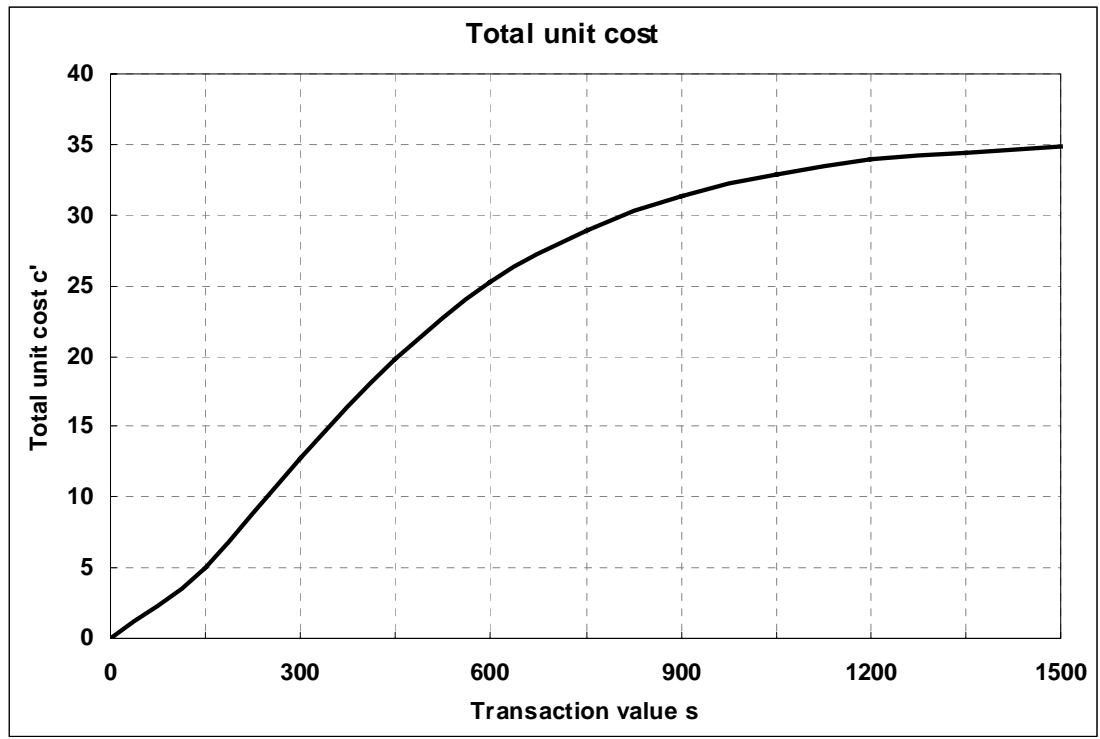

*Figure 5: Unit cost curve*

Note that (ignoring the "small size" premium) although the unit cost for very small transaction values appears to be zero in Table 2 and Figure 5, it is in fact $b$, as shown in Figure 1. The reason is that $b$ is very small, amounting to only 0,3% as shown in Table 3.

## 3.2 Heuristic algorithms

There are three potential heuristic methods which may be applied to solving the problem - genetic algorithms, tabu search and simulated annealing. There is a large amount of research on applications amenable to solution by genetic algorithms and, to a lesser extent, tabu search, and there is readily-available and easy-to-use commercial software to implement these methods. The field of simulated annealing is relatively sparse in comparison, as is the range of software available. Since the development of the optimisation model is intended to be of practical use to practitioners, it was decided to investigate only the performance of genetic algorithms and tabu search.

### 3.2.1 Genetic algorithms

Genetic algorithms (GAs) are adaptive methods which may be used to solve search and optimisation problems. They are based on the genetic processes of biological organisms. Over many generations, natural populations evolve according to the principles of natural selection and "survival of the fittest". By mimicking this process, genetic algorithms are able to "evolve" solutions to real world problems, if they have been suitably encoded. The basic principles of GAs were first laid down rigorously by Holland [24].

GAs work with a *population of individuals*, each representing a possible solution to a given problem. Each individual is assigned a *fitness score* according to how good a solution to the problem it is. The highly-fit individuals are given opportunities to *reproduce*, by *cross breeding* with other individuals in the population. This produces new individuals as *offspring*, which share some features taken from each *parent*. The least fit members of the population of solutions are less likely to be selected for reproduction, and so *die out*.

A whole new population of possible solutions is thus produced by selecting the best individuals from the current *generation*, and mating them to produce a new set of individuals. This new generation contains a higher proportion of the characteristics possessed by the good members of the previous generation. In this way, over many generations, good characteristics are spread throughout the population. By favouring the mating of the more fit individuals, the most promising areas of the search space are explored. If the GA has been designed well, the population will *converge* to an optimal solution to the problem.

A more detailed description of genetic algorithms and their implementation is provided in Appendix I.

### 3.2.2 Tabu search

Tabu search is based on the premise that intelligent problem-solving requires incorporation of adaptive memory and is also a global search technique in that it provides means for escaping from local minima.

In TS, a finite list of forbidden moves called the *tabu list* is maintained. At any given iteration, if the current solution is $x$, its neighborhood $N(x)$ is searched aggressively to yield the point $x'$ which is the best neighbor such that it is not on the tabu list. Often, to reduce complexity, instead of searching all the points in $N(x)$, a subset of these points called the *candidate list* is considered at each step and its size may be varied as the search proceeds. As each new solution $x'$ is generated, it is added to the tabu list and the oldest member of the tabu list is removed. Thus the tabu list inhibits *cycling* by disallowing the repetition of moves within a finite number of steps, as it effectively prevents cycling for cycles shorter than the length of the tabu list. This, along with the acceptance of higher-cost moves, prevents entrapment in local minima.

It may also be desirable to include in the tabu list *attributes* of moves rather than the points themselves. Each entry in the list may thus stand for a whole set of points sharing the attribute. In this case, it is possible to allow certain solutions to be acceptable even if they are in the tabu list by using *aspiration criteria* . For example, one such criterion is satisfied if the point has a cost that is lower than the current lowest cost evaluation. If a neighborhood is exhausted, or if the generated solutions are not acceptable, it is possible to incorporate into the search the ability to jump to a different part of the search space (this is referred to as *diversification*). One may also include the ability to focus the search on solutions which share certain desirable characteristic (*intensification*) by performing pattern recognition on the points that have shown low function evaluations.

```
Initialise
    Identify initial Solution
    Create empty TabuList
    Set BestSolution = Solution
    Define TerminationConditions
done = FALSE
Repeat
    if value of Solution > value of BestSolution then
    BestSolution = Solution
    if no TerminationConditions have been met then begin
        add Solution to TabuList
        if TabuList is full then
            delete oldest entry from TabuList
        find NewSolution by some transformation on Solution
        if no NewSolution was found or
        if no improved NewSolution was found for a long time then
            generate NewSolution at random
    if NewSolution not on TabuList then
            Solution = NewSolution
    end
    else
        done = TRUE
until done = TRUE
```

*Figure 6: Conceptual tabu search algorithm.*

Tabu search is a *metaheuristic* technique, and it must be adapted to the problem for it to be efficient. The choice of moves that generate the neighborhood of a point is problem-specific. Different implementations can be generated by varying the definition and structure of the tabu list, the aspiration criteria, the size of the candidate list, and the intensification and diversification procedures.

TS has been applied successfully to hard problems generally and portfolio optimisation specifically and has been shown to be broadly comparable in performance to GA (see e.g. [30], [21] respectively).

A more detailed description of tabu search and its implementation is provided in Appendix II.

# 4. Results and discussion

## 4.1 Cardinality-unconstrained case

### 4.1.1 Input parameters

The input parameters to the model are shown in Table 4.

| Input parameters | | | | |
|---|---|---|---|---|
| Parameters | Units | | Inputs | Comment |
| Risk-free rate | Fraction | $r_f =$ | 0.130 | 90-d TB rate |
| Market SD | Fraction | $\sigma_m =$ | 0.276 | Measured |
| Risk aversion parameter | Fraction | $w =$ | 0.998 | Various |
| Floor constraint | Fraction | $a =$ | 0.02 | Interviews |
| Ceiling constraint | Fraction | $b =$ | 0.15 | Interviews |
| Asset tradability | Rm/month | $t =$ | 200 | Top 100 stocks average |
| Portfolio size | Rm | $V =$ | 300 | Interviews |
| Include costs? | Binary | Toggle = | 1 | Yes=1, No=0 |
| No. of assets in universe | | | 10 | Interviews |
| Portfolio assets range | Allowed range: | 7 | 50 | Equation (20) |
| Cardinality constraint | Maximum assets: | $K =$ | 10 | OK |

*Table 4: Ten-stock portfolio optimisation parameters*

The risk-free rate used is the current 90-day treasury bill (TB) rate.

In all cases betas and the variance of regression errors have been measured using monthly (month-end) data over the past three years, which is the generally-accepted time period used in practice, i.e. using 36 data points. On this basis, for the JSE all-share index (Alsi) $\sigma_m = 0{,}276$. The risk-aversion parameter $w$ is the variable which is varied to generate the efficient frontiers.

The floor constraint generally used in practice ranges from 1% to 3% and a floor constraint of 2% has been used in the 10-asset-class case, which will be considered first.

In some cases the ceiling constraint is determined legally, for example unit trusts and pension funds are generally restricted to holding a maximum of 5% and 10% respectively in any one stock. A higher ceiling of 15% has been used, however, since sometimes exposure to more than one carefully-selected stock can effectively synthesise an effectively larger holding. These two constraints imply that the number of assets must lie between seven and 50; the cardinality constraint will have to fall within this range. Portfolio managers generally like to keep the number of stocks in a portfolio below 40, keeping in mind that market risk is diversified away with 20-25 stocks.

The universe of JSE-listed shares from which portfolios are created is generally the top hundred stocks in terms of market capitalisation. This universe therefore comprises the Alsi 40 index and the Midcap index (which consist of the next 60 stocks).

The average trade of this universe of the top hundred stocks is presently around R340m/month per stock, but is quite skewed towards the top end; ignoring the top 10 stocks brings the tradability down to R200m/month, and this lower figure has been used. The average portfolio's size in the industry is in the order of R200m-R300m.

The cost function's parameters shown have been determined as described in Section 3.1.

To avoid the problem where the model returns nonzero but insignificantly small asset weightings, an asset is only counted if its weighting exceeds 0,1%, i.e. $z_i = 1$ if $x_i >$ 0,1% in equation (18). Note that this modification is applied only in the calculation of cardinality.

## 4.1.2 Effect of floor and ceiling constraints

In the following three sections a portfolio consisting of 10 stocks has been used to examine the broad effects of various constraints. The stocks selected have been fairly "similar" in terms of returns and variance since a stock with an excessive return or risk level would tend to distort the results in this relatively small portfolio. The portfolio assets' characteristics and other data are shown in Table 5.

| \multicolumn{8}{c}{**Portfolio model**} |||||||||
|---|---|---|---|---|---|---|---|---|
| Asset No. | Asset name | Weight | Number of assets | Forecast return | Total costs | Excess return less costs | Beta (vs Alsi) | Forecast variance |
| i | | $x_i$ | $z_i$ | $r_i$ | $c(x)$ | $R_i$ | $\beta_i$ | $\sigma_{e_i}$ |
| | | (frac) | | (%) | (%) | (%) | (x) | (frac) |
| 1 | A | 0.10 | 1 | 27 | 1.1 | 13.0 | 0.95 | 0.290 |
| 2 | B | 0.10 | 1 | 29 | 1.1 | 15.0 | 0.97 | 0.300 |
| 3 | C | 0.10 | 1 | 35 | 1.1 | 20.9 | 1.12 | 0.245 |
| 4 | D | 0.10 | 1 | 27 | 1.1 | 12.7 | 0.94 | 0.240 |
| 5 | E | 0.10 | 1 | 33 | 1.1 | 19.0 | 1.10 | 0.230 |
| 6 | F | 0.10 | 1 | 40 | 1.1 | 25.5 | 1.00 | 0.213 |
| 7 | G | 0.10 | 1 | 31 | 1.1 | 16.5 | 1.01 | 0.300 |
| 8 | H | 0.10 | 1 | 39 | 1.1 | 24.9 | 1.11 | 0.200 |
| 9 | I | 0.10 | 1 | 35 | 1.1 | 20.7 | 1.16 | 0.195 |
| 10 | J | 0.10 | 1 | 31 | 1.1 | 16.9 | 1.03 | 0.251 |
| Portfolio sum/average | | 1.00 | 10 | 32.6 | | **18.5** | 1.04 | **0.30** |
| Cost function parameters: | | a 30.45 | k 100.00 | c 1.246 | | **Objective function** | | **-0.297** |

*Table 5: Portfolio model structure*

Both floor and ceiling constraints will clearly have a negative impact on a portfolio, for the following reasons.

Floor constraints will force an exposure to every asset, including those with very poor returns, thus reducing the portfolio's return.

For low levels of risk aversion the portfolio will normally tend to consist of only one or two assets, i.e. those with the highest returns. The ceiling constraint, however, will make the high optimal level of exposure to these assets impossible, and again force an exposure to lower-returning assets, which will reduce the portfolio's return.

The effect of floor and ceiling constraints on the efficient frontier was calculated and is shown in Figure 7.

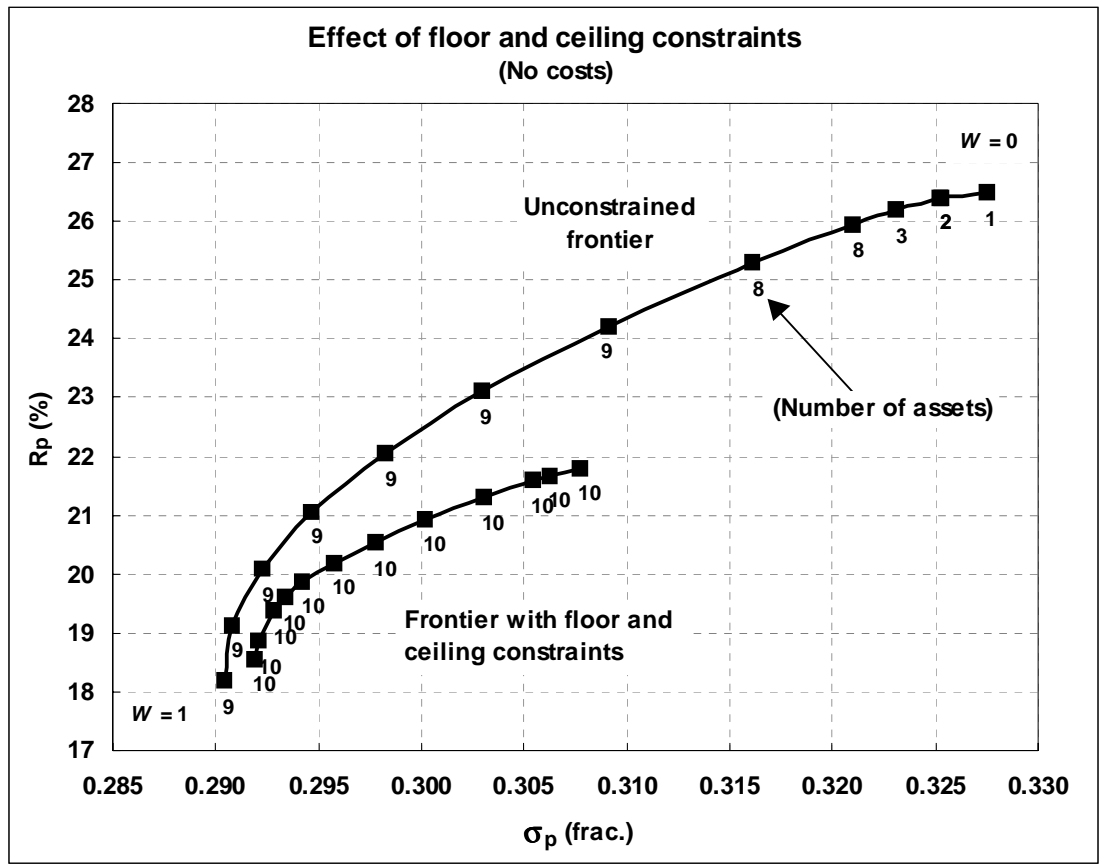

*Figure 7: Effect of floor and ceiling constraints*

With no constraints the highest-returning portfolio consists only of one asset, yielding a return of 26,5%. With the floor and ceiling constraints the highest possible return of the (now 10-asset) portfolio is only 21,8%, although the greater number of assets has also reduced the portfolio's risk from 0,327 to 0,307. However, the lowest-risk constrained portfolio has a higher variance than that of the unconstrained portfolio, since the constraints also interfere with the optimal weights for risk reduction. The constrained portfolio is therefore completely dominated by the unconstrained portfolio.

The impact on risk in this particular example is, however, smaller than the impact on returns. This may not necessarily be true in general; the relative impact is dependent on many variables, including the shape of the efficient frontier (which in turn is dependent on the absolute levels of forecast returns and forecast risks for all its constituents, as well as their cross-correlations) and the absolute magnitudes of the floor and ceiling constraints.

### 4.1.3 Effect of costs

The impact of costs on the portfolio, without any floor or ceiling constraints, is shown in Figure 8.

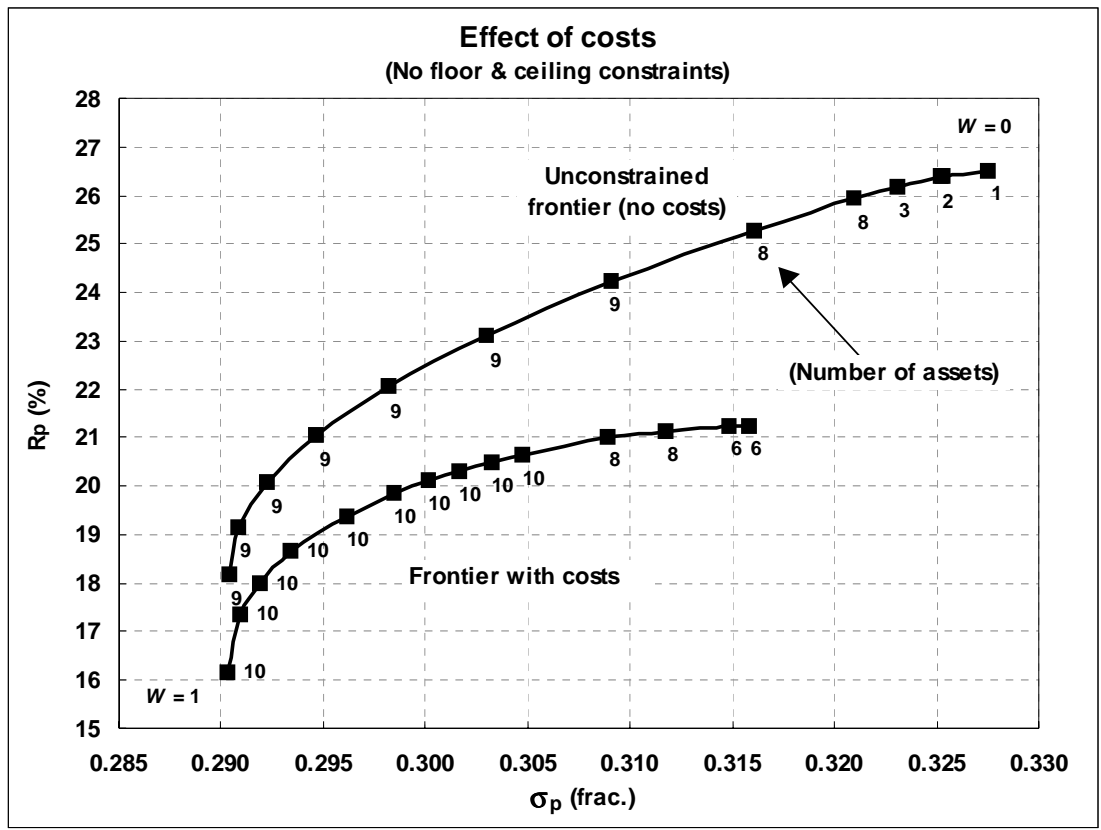

*Figure 8: Effect of costs*

Without costs or floor and ceiling constraints, the highest-returning portfolio again consists only of one asset. However, the large size of the order required will result in a high transaction cost since costs increase exponentially with order size. If this cost is of a magnitude comparable to the forecast returns, the portfolio will tend to diversify into more assets in order to reduce total transaction costs and their adverse impact on returns. This results in the following important insight:

***Transaction costs will tend to diversify portfolios.***

This is shown quite strikingly in Figure 8, where the least risk-averse, highest-returning portfolio consists of as many as six assets instead of one, purely as a result of attempting to reduce costs. The stocks selected tend to be those with the highest returns.

The most risk-averse portfolio has approximately the same risk as the cost-free risk-averse portfolio, since both are quite fully diversified in terms of number of stocks. The stocks selected at this end of the frontier tend to be those with the lowest betas.

However, the return of the cost-laden portfolio is lower, by the amount of the total transaction costs incurred. As with floor and ceiling constraints, the cost-laden portfolio is completely dominated by the cost-free portfolio.

### 4.1.4 Combined effect of floor and ceiling constraints and costs

The impact on the portfolio of both floor and ceiling constraints as well as costs is shown in Figure 9 on page 34.

The negative impact on the constrained and cost-laden portfolio is cumulative. The three frontiers are shown in Figure 10 on page 34 and their differences are summarised in Table 6 on page 35.

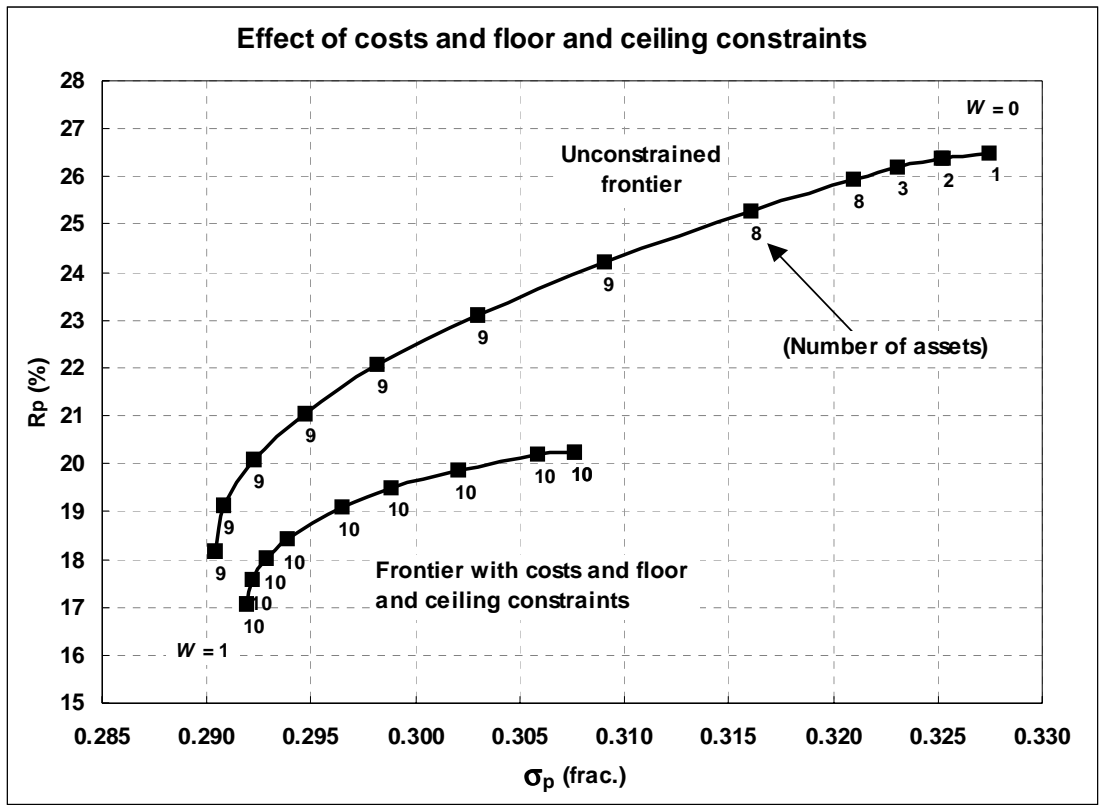

*Figure 9: Combined effect of constraints and costs*

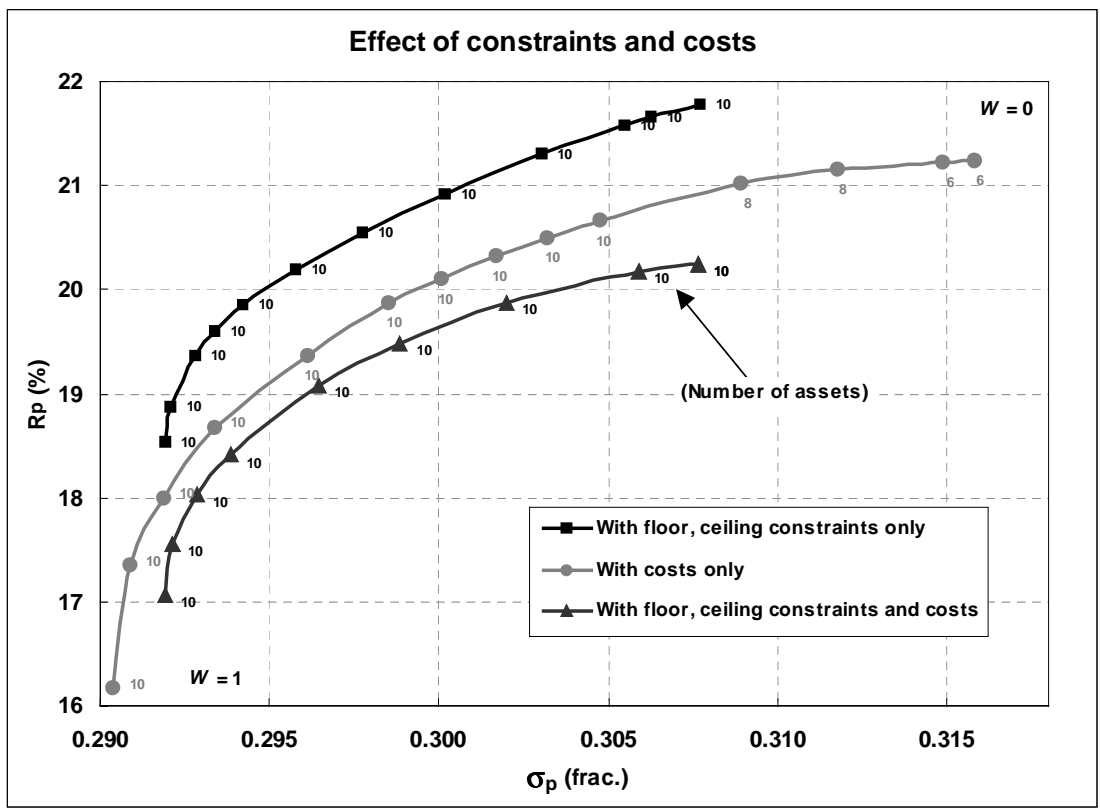

*Figure 10: Summary of efficient frontiers*

| Summary |||| 
|---|---|---|---|
| Constraints | Portfolio characteristic | Highest-return portfolio | Lowest-risk portfolio |
| No constraints | Return (%) | 26.50 | 18.18 |
|  | Risk (%) | 0.328 | 0.290 |
|  | Stocks no. | 1 | 8 |
| Floor & ceiling | Return (%) | 21.78 | 18.53 |
|  | Risk (%) | 0.308 | 0.292 |
|  | Stocks no. | 10 | 10 |
| Costs | Return (%) | 21.24 | 16.16 |
|  | Risk (%) | 0.316 | 0.290 |
|  | Stocks no. | 6 | 9 |
| Floor & ceiling and costs | Return (%) | 20.24 | 17.07 |
|  | Risk (%) | 0.308 | 0.292 |
|  | Stocks no. | 10 | 10 |
| Difference vs. unconstrained | Return (%) | -6.26 | -1.11 |
|  | Risk (%) | -0.020 | 0.001 |
|  | Stocks no. | 9 | 2 |

*Table 6: Summary of constraint and cost effects*

The imposition of both constraints and costs reduces returns for both the highest-return portfolio and the lowest-risk portfolio, although the effect is more marked in the case of the highest-return portfolio, where the decline of 6,26% is almost a quarter of the portfolio's total return.

Risk for the highest-return portfolio is reduced by the introduction of constraints and costs, since they tend to diversify the portfolio. However, for the lowest-risk portfolio, floor and ceiling constraints will increase risk since they force an exposure to high-risk assets which could otherwise be avoided.

***An interesting result is that this may in some cases also be accompanied by a corresponding increase in return.***

This is shown in Table 6. The impact of costs is thus ambiguous and may either increase or reduce risk, depending on the degree of diversification required for cost reduction. Risk is increased in this example, where there are both constraints and costs.

*The number of assets in the portfolio invariably increases as a result of constraints, costs and their combination.*

### 4.1.5 The need for cardinality constraints

Note that in the 100-stock portfolio which is to be optimised, if the desired floor constraint is 1%, often used in practice, then all the asset class weights $x_i$ cannot be anything other than 1% - this single constraint has determined the portfolio structure! In order to "optimise" the portfolio either a lower floor constraint would be required, which would not satisfy the actual minimum level desired, or the number of stocks in the universe needs to be reduced from 100, which could be an undesirable contraction of the investable universe.

The manner in which the floor constraint was implemented in the previous example meant that *all* stocks in the investable universe were included in the portfolio at that minimum weighting, with no stocks having a zero weighting, no matter how unattractive. What the floor constraint actually means in practice is that if a stock is selected, it will be included in the portfolio at above the floor weighting; if not, its weighting will be zero. Cardinality constraints are therefore disjunctive in nature.

This underlines the fact that the application of cardinality constraints is essential in any portfolio optimisation that claims to be realistic.

## 4.2 Cardinality-constrained case

### 4.2.1 Testing the heuristic methods

For the real-world cardinality-constrained 100-stock portfolio with both floor and ceiling constraints and (nonlinear) costs there is no method of calculating the exact efficient frontier any more because of the mixed-integer constraints and the size of the problem, and hence no way of benchmarking the heuristic methods against the exact solution. Therefore, to test initially the effectiveness of the heuristic methods and establish their credentials, they are first used to find the efficient frontier without any cardinality constraints. Unless they are able to do this with a reasonable amount of accuracy and within a reasonable time, they are unlikely to find the cardinality-constrained efficient frontier successfully. The model is easily set to the cardinality-unconstrained case by setting $K = N$. The floor, ceiling and cost constraints are, however, retained in this test case.

The test set of data is presented in Appendix III. The portfolio consists of the top 100 stocks by market capitalisation on the Johannesburg Stock Exchange (JSE). Forecast returns are compound estimates over the next two years and variances are the actual historical levels as measured over the past 36 months. Stocks with histories of less than 20 months are flagged and judgemental estimates have to be used in some of these cases. To avoid the problem of the constraints determining the portfolio structure, as mentioned in the Section 4.1.5, the floor constraint was set at 0,5% for this test case.

The portfolio parameters used are shown in Table 7.

| Input parameters | | | | |
|---|---|---|---|---|
| Parameters | Units | Inputs | | Comment |
| Risk-free rate | Fraction | $r_f =$ | 0.130 | 90-d TB rate |
| Market SD | Fraction | $\sigma_m =$ | 0.276 | Measured |
| Risk aversion parameter | Fraction | $w =$ | 0.000 | Various |
| Floor constraint | Fraction | $a =$ | 0.005 | Interviews |
| Ceiling constraint | Fraction | $b =$ | 0.150 | Interviews |
| Asset tradability | Rm/month | $t =$ | 200 | Top 100 stocks average |
| Portfolio size | Rm | $V =$ | 300 | Interviews |
| Include costs? | Binary | Toggle = | 1 | Yes=1, No=0 |
| No. of assets in universe | | | 100 | Interviews |
| Portfolio assets range | Allowed range: | 7 | 200 | Equation (20) |
| Cardinality constraint | Maximum assets: | $K =$ | 100 | OK |

*Table 7: Hundred-stock portfolio optimisation parameters*

The cardinality-unconstrained efficient frontier can be found using a commercial package as such as LINGO, which can solve nonlinear problems involving both continuous and binary variables using branch-and-bound methods, or Excel's Solver in nonlinear mode. The parameters used in the optimisation by Solver are shown in Table 8.

| Quadratic programming problem | |
|---|---|
| **Optimisation parameters** | |
| Time limit | 300 sec |
| Iteration limit | 300 |
| Precision | 0.00001 |
| Tolerance | 3% |
| Convergence | 0.0001 |
| Model | Nonlinear |
| Variables | Nonnegative |
| Scaling | Automatic |
| Initial estimates | Quadratic extrapolation |
| Partial derivatives | Central differencing |
| Search direction | Quasi-Newton |
| Average solution time | 215 sec |
| Average iterations | 119 |

*Table 8: QP parameter settings*

The *precision* with which variables such as asset class weights were required to meet the constraints or targets was 0,001%, while *tolerance* refers to integer constraints, the only one being that the asset class weights must sum to one. When the objective function changes by less than the *convergence* amount the iteration stops. The optimisation is speeded up by specifying that all input variables (the asset class weights) are nonnegative - in other words short sales are not allowed, as discussed in Section 1.3. *Scaling* is required since the magnitude of the forecast returns can be as much as three orders of magnitude larger than the forecast variance. Quadratic extrapolation is used since the problem can be highly nonlinear.

On a single 500MHz processor and 196MB of RAM under Windows NT 4.0 each point on the frontier took approximately 120 iterations and three to four minutes to calculate. The resultant efficient frontier is presented in Figure 11 on page 39.

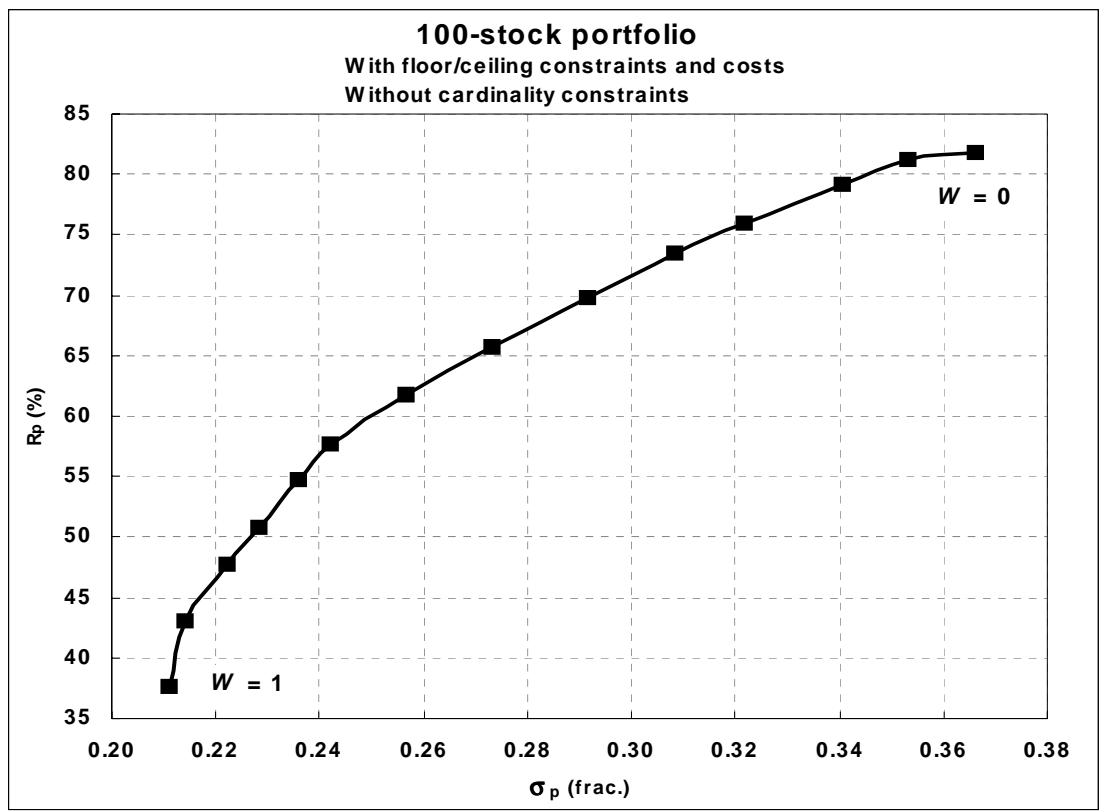

*Figure 11: Hundred-stock cardinality-unconstrained efficient frontier*

### 4.2.2 Testing on the cardinality-unconstrained case

In constructing efficient frontiers, the value of $w$ was always increased sequentially from zero to one and the asset class weights were not reset for each run. The best solution found for any value of $w$ should then be a good starting point for the next value of $w$, thus shortening optimisation times.

Starting with equal asset weights, solution times tend to decrease with increasing $w$, since a high $w$ requires a more diversified portfolio than for a low $w$, and this starting point provides a closer approximation to this solution.

Where the heuristic methods used random-number generators, the initial seed was never randomised but set to 999 in order to assist reproducibility.

**Tabu/scatter search**

The software used was the Optquest module of Decisioneering Inc.'s Crystal Ball, an optimisation and simulation product. The optimisation parameters used are shown in Table 9.

| Tabu/scatter search Optimisation parameters | |
|---|---|
| Time limit | 60 min |
| Iteration limit | 1000 |
| Population size | 20 |
| Tolerance range multiplier | 0.005 |
| Variable type | Discrete |
| Step size | 0.001 |
| NN accelerator | Off |
| Gradient search | On |
| Taguchi design | Off |
| Max. trials | 1 |
| Burst amount | 1000 |

*Table 9: TS parameter settings*

The recommended number of iterations for a problem with 100 decision variables is at least 5000. However, this would have resulted in impractically long runs and the effective limit used was a 60-minute run time.

*Population* refers to the number of solution sets in the tabu scheme. The population is selected by comparing the time of the first iteration to the time limit for the search. For fast iterations and long time limits it is capped at 100. For slower optimisations, a population size of 15% of the estimated total number of iterations is used, with a lower limit equal to the number of decision variables. For the 100-stock portfolio the population was therefore effectively always 100.

To maximise speed the decision variables (ie. the asset class weights) were assumed to be *discrete* rather than continuous, with an implied precision or *step size* of 0.1% being sufficient for practical purposes. The *tolerance range multiplier* is used to distinguish equivalent solutions and reject one of them. Since the maximum asset class weight is 15%, the value $(0,005)(0,15) = 0,00075$ was used, which is below the step size.

*Trials* refers to a stochastic mode; this was not used as the model uses deterministic inputs. *Burst amount* is an information communication parameter. No *stopping rule* was used initially.

A typical convergence path for this method is shown in Figure 12.

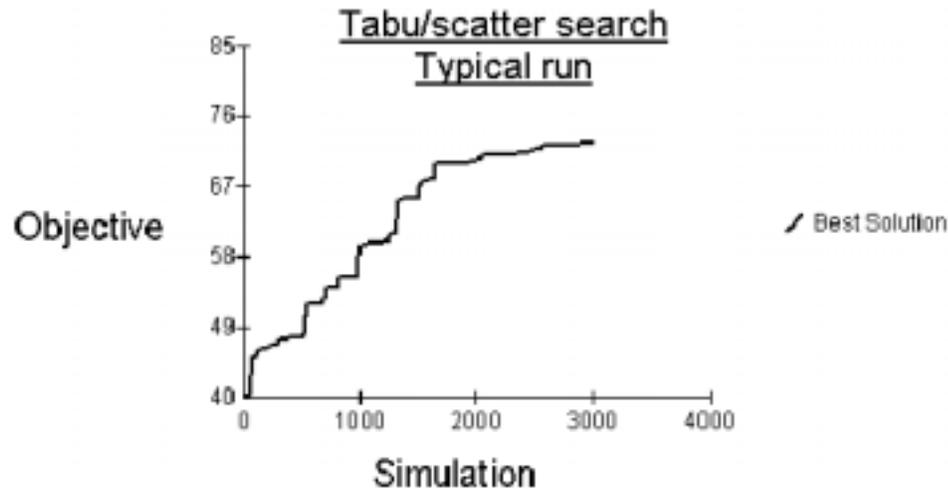

*Figure 12: TS typical convergence*

**Genetic algorithm**

The software used was Evolver Professional, by Palisade Corporation. The optimisation parameters are shown in Table 10 on page 42.

A *population* of 30-100 is usually used, with larger populations being necessary for larger problems. A larger population takes longer to converge to a solution but is more likely to find the global optimum because of its more diverse gene pool.

The *mutation rate* is increased when the population ages to an almost homogenous state and no new solutions have been found for a few hundred trials. Mutation provides a small amount of random search, and helps ensure that no point in the search has a zero probability of being examined.

The portfolio problem is unable to benefit from certain specifically-tailored *genetic operators*, so the default values for these were used. No *stopping rule* was used initially. All other settings were left at their default values.

| GA search | |
|---|---|
| **Optimisation parameters** | |
| Time limit | 30 min |
| Iteration limit | 50 000 |
| Method | Budget |
| Population size | 50 |
| Crossover rate | auto (0.06 usually) |
| Mutation rate | 0.5 |
| Genetic operators: | |
|   parent selection | Default |
|   mutation | Default |
|   crossover | Default |
|   backtrack | Default |
| Random seed | 999 |
| Stop on change | <5% in 1000 trials |
| Average solution time | 11 min |
| Average iterations | 18 419 |

*Table 10: GA parameter settings*

A typical convergence path for this method is illustrated in Figure 13. The upper line represents the best solution in the population and the lower line the average solution.

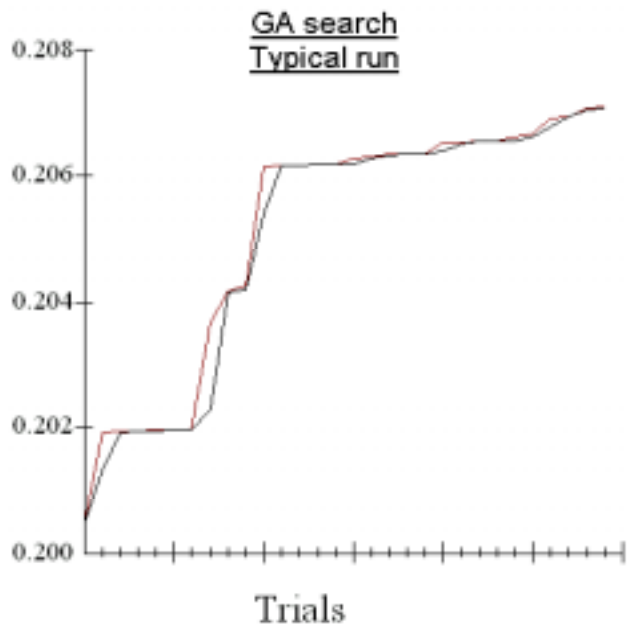

*Figure 13: GA typical convergence*

**Results**

The efficient frontiers generated by the two methods are shown in Figure 14 and the value of the objective function for various values of the risk-aversion parameter *w* are presented in Figure 15 on page 44. Note that the *x*-axis scales in the following graphs do not have equal increments throughout.

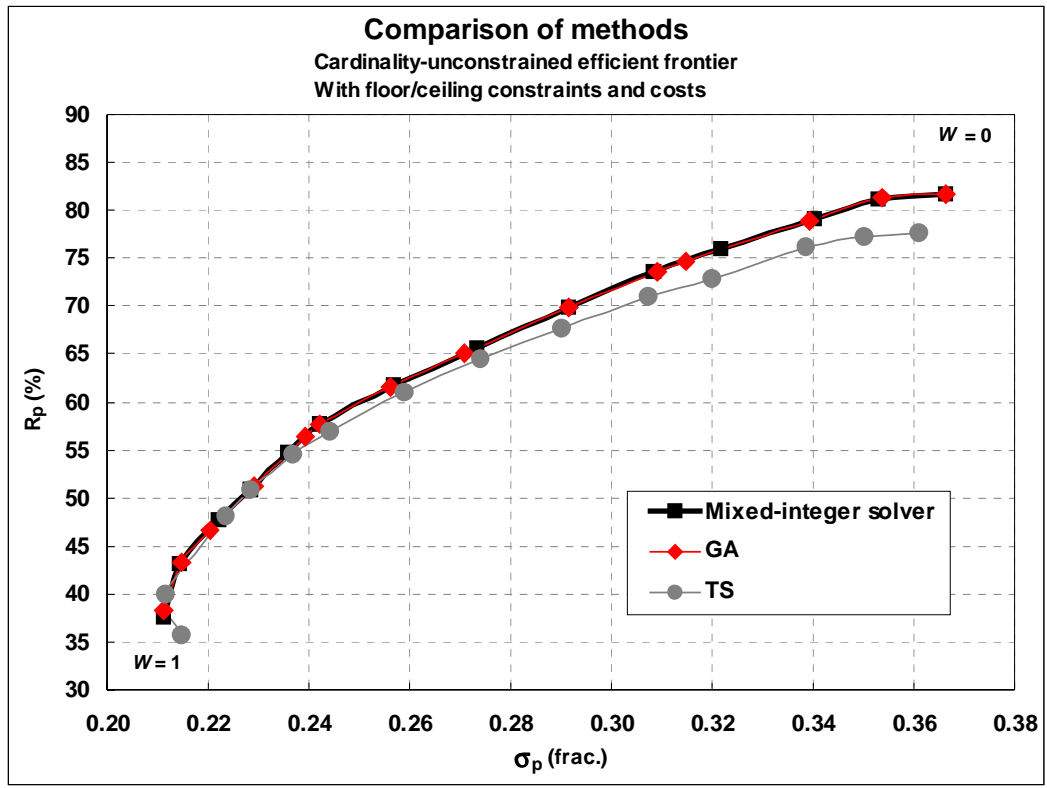

*Figure 14: Heuristic test frontiers*

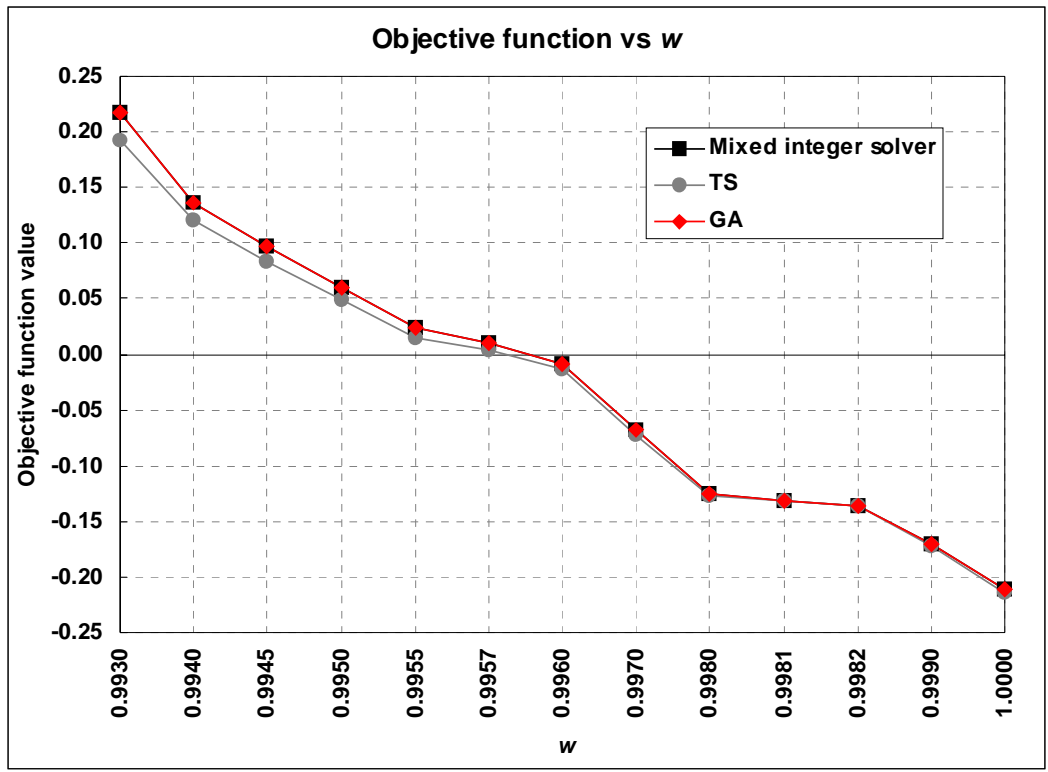

*Figure 15: Objective function values*

The comparative results of the two heuristic methods for the cardinality-unconstrained case are summarised in Table 11. The complete set of data is provided in Appendix IV.

| | | Comparison of heuristic tests | | | | | |
|---|---|---|---|---|---|---|---|
| | | Absolute error | | | Solution time (min) | Best trial (no.) | Total trials (no.) |
| | | Return (%) | Risk (%) | Objective (%) | | | |
| TS | Median | 3.30 | 0.54 | 8.56 | 60 | 371 | 723 |
| | Standard deviation | 2.06 | 0.46 | 20.62 | 6 | 271 | 175 |
| | Mean | 3.06 | 0.73 | 16.48 | **59** | 379 | 736 |
| | Combined mean | **1.89** | | - | - | - | - |
| GA | Median | 0.36 | 0.24 | 0.07 | 8 | 14819 | 14819 |
| | Standard deviation | 0.95 | 0.62 | 0.09 | 7 | 10656 | 10656 |
| | Mean | 0.85 | 0.51 | 0.08 | **11** | 18419 | 18419 |
| | Combined mean | **0.68** | | - | - | - | - |

*Table 11: Heuristic test performance*

While the tabu/scatter heuristic worked well on small (around 20 assets) problems, its rate of convergence to the solution slowed dramatically when the problem size increased to 100 assets. A stopping rule of 60 minutes was therefore implemented. The method's error is thus a reflection of not being given enough time to find a sufficiently accurate solution and not necessarily an inherent inability to eventually find that solution.

The mean error in the efficient frontier calculated by TS was almost 16,5% after an average 59 minutes. Even after this long calculation time, some points had errors of over 60%, resulting in a large standard deviation of 21%. There were larger errors at the upper end of the frontier ($w = 0$) as the solution at this end of the frontier usually consists of only a few stocks, which is further from the initial solution used of an equal-weighted portfolio than the highly-diversified situation at the other end of the frontier, as discussed previously.

The "distance" of the calculated frontier from the benchmark frontier was measured in a rudimentary way by the arithmetic average of the (absolute) percentage errors in both return and risk. This combined error was 1,9% for the TS heuristic.

In comparison, the genetic algorithm provided a mean error of only 0,08% in an average calculation time of only 11 minutes. The maximum single error was only 0,38% and the standard deviation relatively low at 0,09%. The mean absolute error for both estimated returns and risk was 0,68%. Ignoring the time factor, the accuracy of GA was over 200 times better than TS for the objective function value and nearly three times better for the combined return and risk measure. It is interesting to note that the standard deviation of the errors for both returns and risk was less than the mean for TS but larger than the mean for GA, giving the latter a larger coefficient of variation.

For the TS heuristic the median error of the objective function is significantly smaller than the mean error, which implies a skew error distribution with a higher probability of large errors. In comparison, the GA's median and mean errors are approximately equal, indicating a symmetric error distribution for this function. However, examining the errors in returns and risk separately, the TS error distribution for risk may be skew, while for GA those for both risk and return could be skew.

However, a fairer indication is provided by combining (absolute) accuracy and time by using their simple product as the performance criterion. Figure 16 shows both methods' performance across the frontier for various values of *w*.

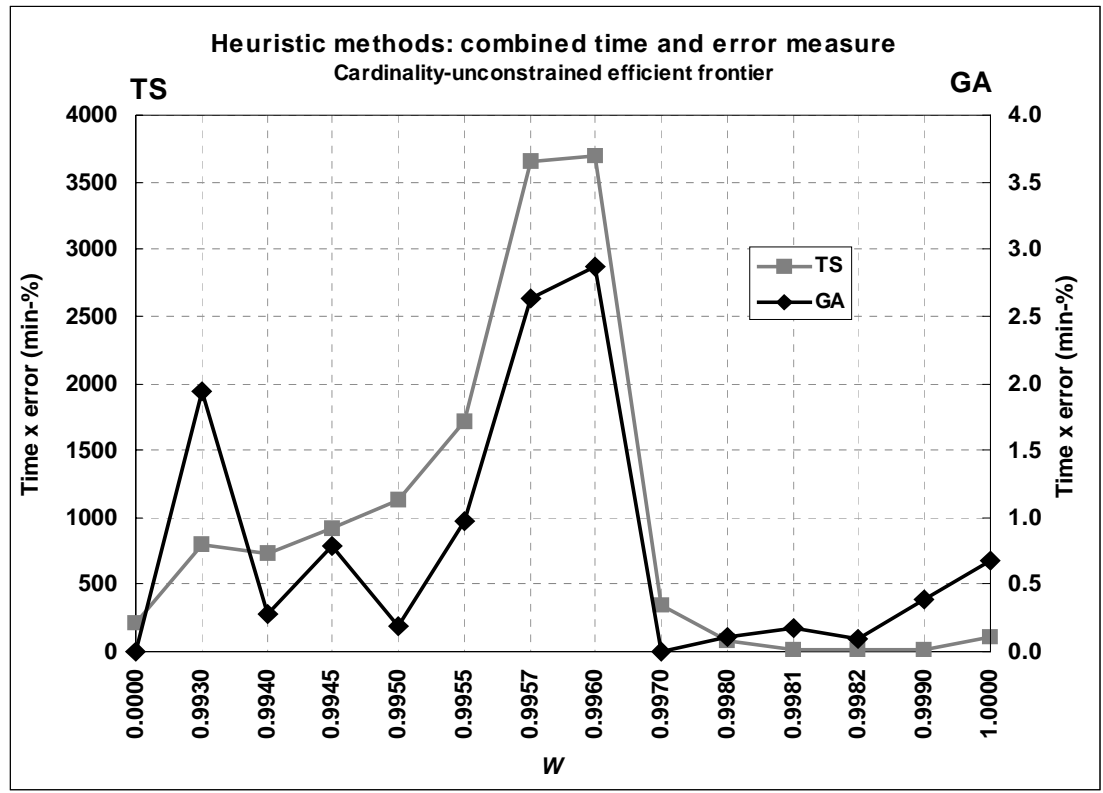

*Figure 16: Efficiency of heuristic methods*

***On the basis of this measure the performance of GA is better than that of TS by approximately three orders of magnitude!***

Interestingly, both methods found the centre portion of the efficient frontier the most difficult to generate. A possible reason is that at the upper end (highest returns, low risk aversion, $w = 0$) the selection of the highest-return stocks is relatively straightforward, and at the lower end (lowest risk, $w = 1$) the strategy is also simple: select the lowest-beta stocks. However, in the central part of the frontier there is a much larger number of combinations of stocks that will result in middle-of-the-road return and risk levels.

For the GA method there was no relationship whatsoever between the accuracy of a point and the number of trials required to achieve this ($r^2 = 0,002$).

The relationship between run time and the number of trials also had a large amount of scatter ($r^2 = 0,46$). This implies that the time required per trial varies, and it was in fact found to decline with increasing $w$, possibly for the reason discussed previously. This is shown in Figure 17.

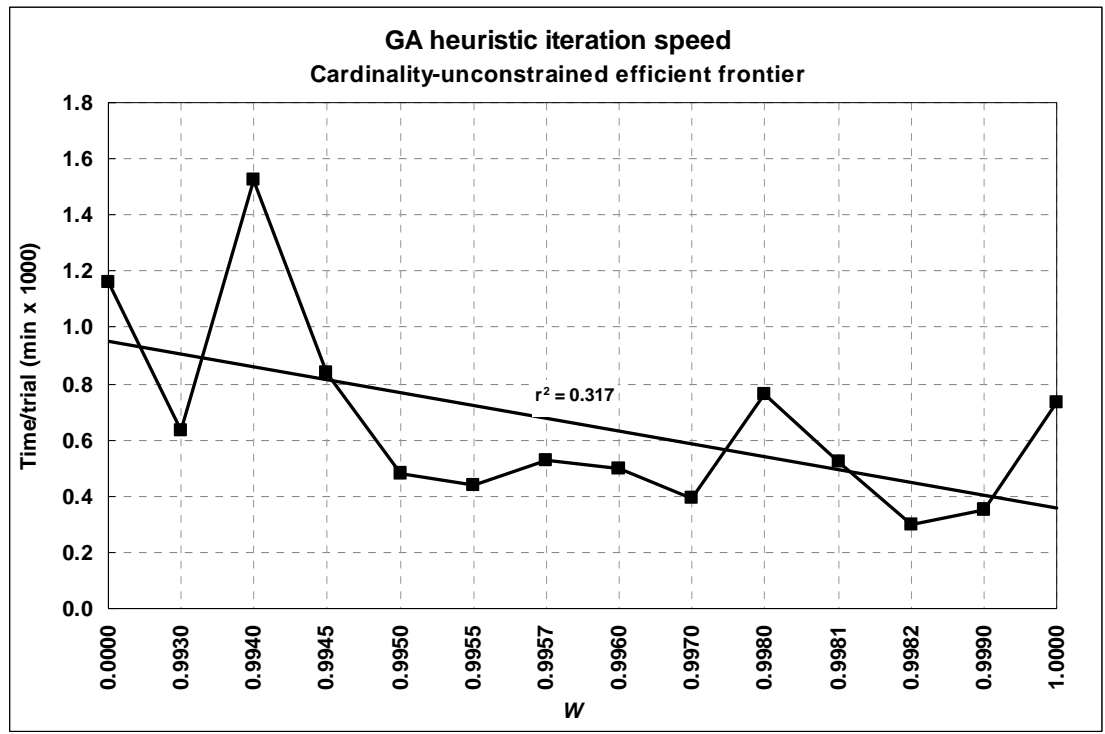

*Figure 17: GA iteration speed across frontier*

These tests indicate that while both heuristics may be used to generate the efficient frontier, for a problem of this size the performance of the GA is orders of magnitude better than that for TS. The GA was able to find solutions arbitrarily close to the correct value, given sufficient (but quite reasonable) calculation times. Also, perhaps more time should be allocated to the points in the central part of the frontier.

It may be noted that since the efficient frontier found by the heuristic methods consists of suboptimal points, it will always be dominated by the true efficient frontier.

***The optimal portfolios generated for the cardinality-constrained case will therefore always be conservative***.

This is because for any selected level of return the indicated risk level will be higher than the true level, while for any selected level of risk the calculated return will be lower than the actual return.

### 4.2.3 Application to the cardinality-constrained case

In the cardinality-constrained case the floor constraint is subsumed into the cardinality count, i.e. if $x_i > 0,5\%$ then $z_i = 1$, otherwise it is zero. The ceiling of 15% is retained, which sets the minimum number of assets at 6,7, i.e. 7. A cardinality constraint of 40 stocks within the 100-stock universe was selected.

Note that the cardinality constraint can just as easily and less restrictively be set to a range, e.g. $K_l \leq K \leq K_u$, where $K_l$, $K_u$ represent the lower and upper limits on the number of assets in the portfolio respectively.

While already slow on large problem instances, the tabu/scatter method is particularly ill suited to the cardinality-constrained case, since it runs an entire optimisation before determining whether the result is cardinality-constraint infeasible. To avoid running these iterations, it must identify the characteristics of solutions likely to be infeasible, which makes the search more complex and can extend the search time by over 50%.

Nevertheless, two attempts were made to find the cardinality-constrained efficient frontier with TS, for $w = 0$ and $w = 1$. In both cases not only was the iteration speed impractically slow, but there were no convincing signs of any probable convergence, as shown in Figures 18 and 19 on page 49.

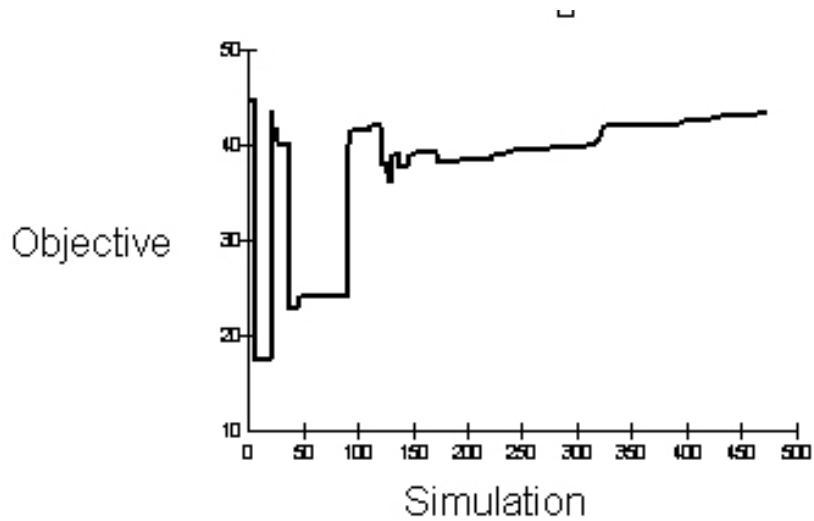

*Figure 18*: Cardinality-constrained TS convergence: $w = 0$

The $w = 0$ optimisation took 3 hours to complete 530 trials. The best objective function value after this time was 44,1 (in comparison, the cardinality-unconstrained optimum was 81,8).

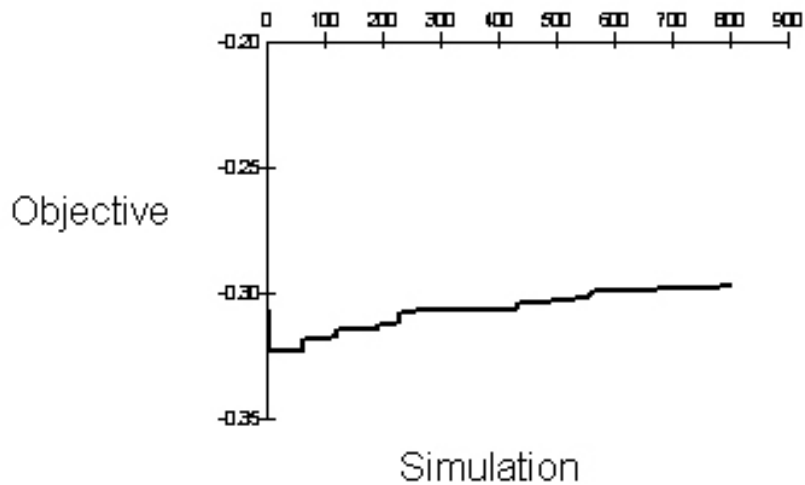

*Figure 19*: Cardinality-constrained TS convergence: $w = 1$

The $w = 1$ optimisation took 3 hours to complete 802 trials. The best objective function value after this time was –0,297 (in comparison the cardinality-unconstrained optimum was 0,211).

*The tabu/scatter search method in this type of implementation is therefore unsuitable for the optimisation of portfolios of this size.*

The introduction of cardinality constraints may result in a discontinuous efficient frontier. The discontinuities imply that there are certain combinations of return and risk which are "undefined" for a rational investor, since an alternative portfolio with both a higher return and lower risk exists. An example from the paper by Chang *et al* [21] is shown in Figure 20.

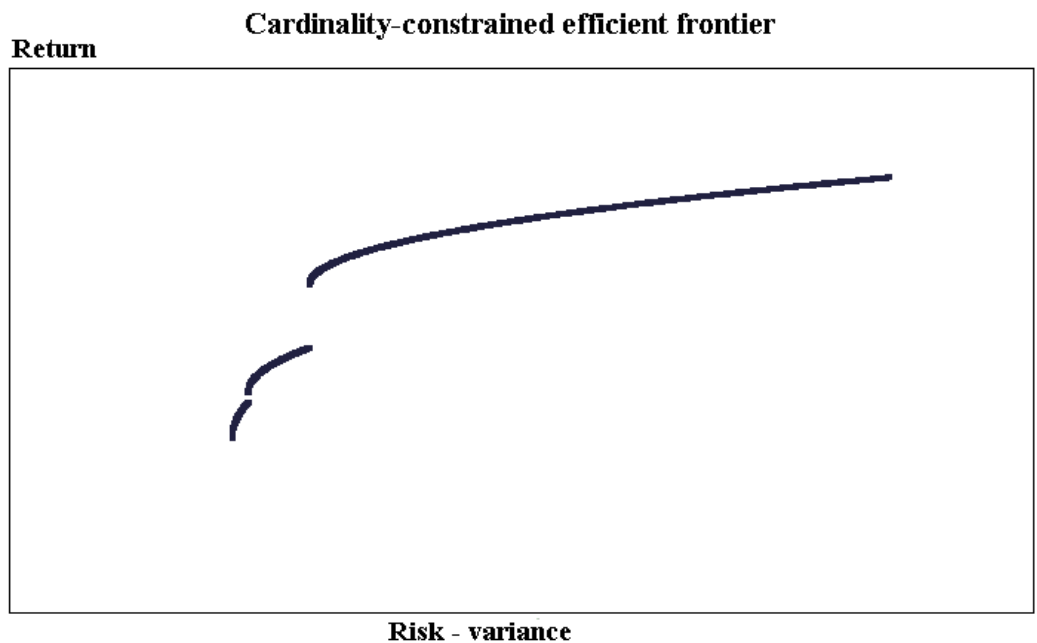

*Figure 20: Typical discontinuous cardinality-constrained frontier*

The cardinality-constrained efficient frontier for $K = 40$ stocks was constructed using 22 different values of *w*. The curve is shown in Figure 21 on page 51 and its values in Table 12 on page 52.

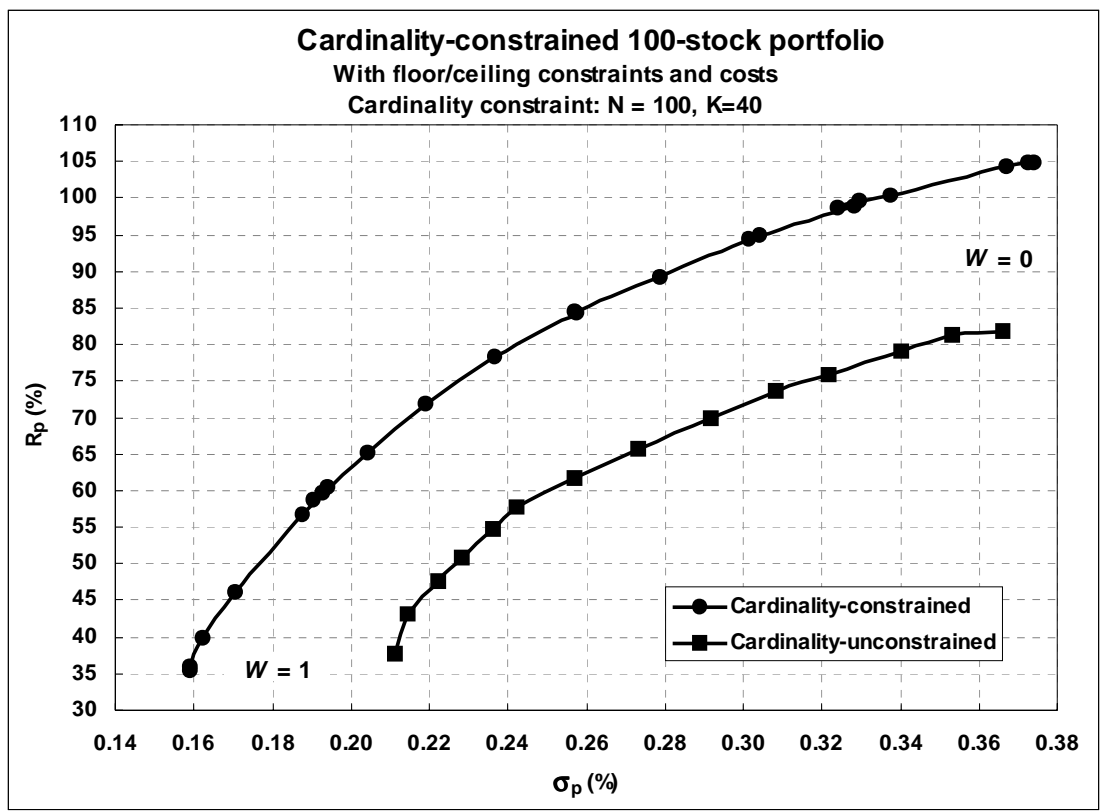

*Figure 21: Cardinality-constrained 100-stock efficient frontier*

There were no signs of any discontinuities in this particular cardinality-constrained efficient frontier.

The computation times were approximately a third of those reported by Chang *et al* for a portfolio of similar size, after making a rough adjustment for processor speed.

It may be noted that even in larger markets with a far larger universe of listed stocks, the universe of portfolio candidates is unlikely to be dramatically larger than the $N = 100$ used, as quantitative or other methods are often used to do the initial filtering

The cardinality-unconstrained efficient frontier is shown on the same graph for comparison. The cardinality-constrained portfolio completely dominates the cardinality-unconstrained portfolio. On average, for the same level of return the cardinality-constrained frontier exhibits risk which is lower by between 0,05 to 0,125 or 5% and 12,5%.

| Cardinality-constrained efficient frontier | | | | | | |
| N = 100, K = 40 | | | | | | |
| w | Return (%) | Risk (%) | Objective function (%) | No. of stocks (no.) | Solution time (min) | Best trial (no.) | Total trials (no.) |
| --- | --- | --- | --- | --- | --- | --- | --- |
| 0.0000 | 104.86 | 0.373 | 104.864 | 40 | 27 | 10 093 | 37 365 |
| 0.5000 | 104.76 | 0.374 | 52.194 | 40 | 39 | 19 620 | 44 321 |
| 0.9900 | 104.37 | 0.367 | 0.680 | 40 | 19 | 7 053 | 25 124 |
| 0.9930 | 100.31 | 0.337 | 0.367 | 40 | 34 | 10 673 | 55 603 |
| 0.9940 | 99.49 | 0.329 | 0.270 | 40 | 36 | 14 022 | 45 231 |
| 0.9945 | 98.51 | 0.324 | 0.220 | 40 | 34 | 14 109 | 51 301 |
| 0.9950 | 98.91 | 0.328 | 0.168 | 40 | 29 | 9 987 | 37 666 |
| 0.9953 | 94.83 | 0.304 | 0.143 | 40 | 25 | 8 657 | 33 035 |
| 0.9955 | 94.40 | 0.302 | 0.121 | 40 | 34 | 11 302 | 43 582 |
| 0.9956 | 89.31 | 0.279 | 0.115 | 40 | 34 | 15 624 | 41 150 |
| 0.9957 | 84.53 | 0.257 | 0.108 | 40 | 55 | 18 410 | 45 008 |
| 0.9960 | 84.30 | 0.258 | 0.081 | 40 | 30 | 15 979 | 35 987 |
| 0.9970 | 78.34 | 0.237 | -0.001 | 40 | 30 | 12 002 | 44 678 |
| 0.9975 | 71.90 | 0.219 | -0.039 | 40 | 30 | 13 312 | 42 342 |
| 0.9980 | 65.08 | 0.204 | -0.074 | 40 | 23 | 11 166 | 37 770 |
| 0.9981 | 59.65 | 0.192 | -0.079 | 40 | 30 | 11 492 | 47 322 |
| 0.9982 | 58.62 | 0.191 | -0.085 | 40 | 28 | 10 714 | 58 202 |
| 0.9981 | 60.47 | 0.194 | -0.079 | 40 | 30 | 13 713 | 40 528 |
| 0.9984 | 56.70 | 0.187 | -0.096 | 40 | 30 | 11 608 | 43 652 |
| 0.9986 | 46.06 | 0.171 | -0.106 | 40 | 36 | 18 003 | 43 756 |
| 0.9988 | 39.84 | 0.162 | -0.114 | 40 | 30 | 15 269 | 38 942 |
| 0.9990 | 36.00 | 0.159 | -0.123 | 40 | 36 | 18 719 | 46 122 |
| 1.0000 | 35.48 | 0.159 | -0.159 | 40 | 30 | 12 116 | 37 544 |

*Table 12: Cardinality-constrained 100-stock efficient frontier*

Conversely, for equal risk levels the cardinality-constrained portfolio produces higher returns, which range from 24% to 30% higher across the efficient frontier. This is substantial, given that the average return of the cardinality-unconstrained portfolio is 63%.

- *Finding the best subset of the universe of stocks rather than optimising the universe itself results in a dramatically better portfolio.*

The ability of the heuristic model to optimise cardinality-constrained portfolios is one of its most powerful features.

The next step in optimising a real-world portfolio is to determine the risk-aversion factor *w*. This is done easily from the original Markowitz theory. The capital market line or capital allocation line (CAL) is drawn from the point representing risk-free T-bills to the efficient frontier; the optimal risky portfolio is represented by the point where the CAL is tangent to the efficient frontier - at this point the CAL has the steepest slope and thus offers the highest return-to-risk ratio. The T-bill point is represented by the risk-free interest rate $r_f$ (which was 10,3% for 90-day T-bills when this study was begun) and effectively zero variance or risk. Most portfolios have a cash component; in some types of fund there is a minimum legal requirement.

The value of *w* at the point of tangency is the risk-aversion parameter which will be used to optimise the cardinality-constrained portfolio. It must be noted, however, that while this value of *w* is optimal, it is not necessarily the *w* an investor would choose if they could hold only the risky portfolio, since it assumes that they can hold a portfolio of both cash and the risky portfolio. In this case the optimal risky portfolio is always used, irrespective of their risk preferences; their risk aversion comes into play not by choosing a different *w*-point on the efficient frontier but in the selection of their desired point on the CAL, i.e. the mixture of optimal risky portfolio and risk-free cash. So for the risky portfolio on its own the investor could have a quite different *w*.

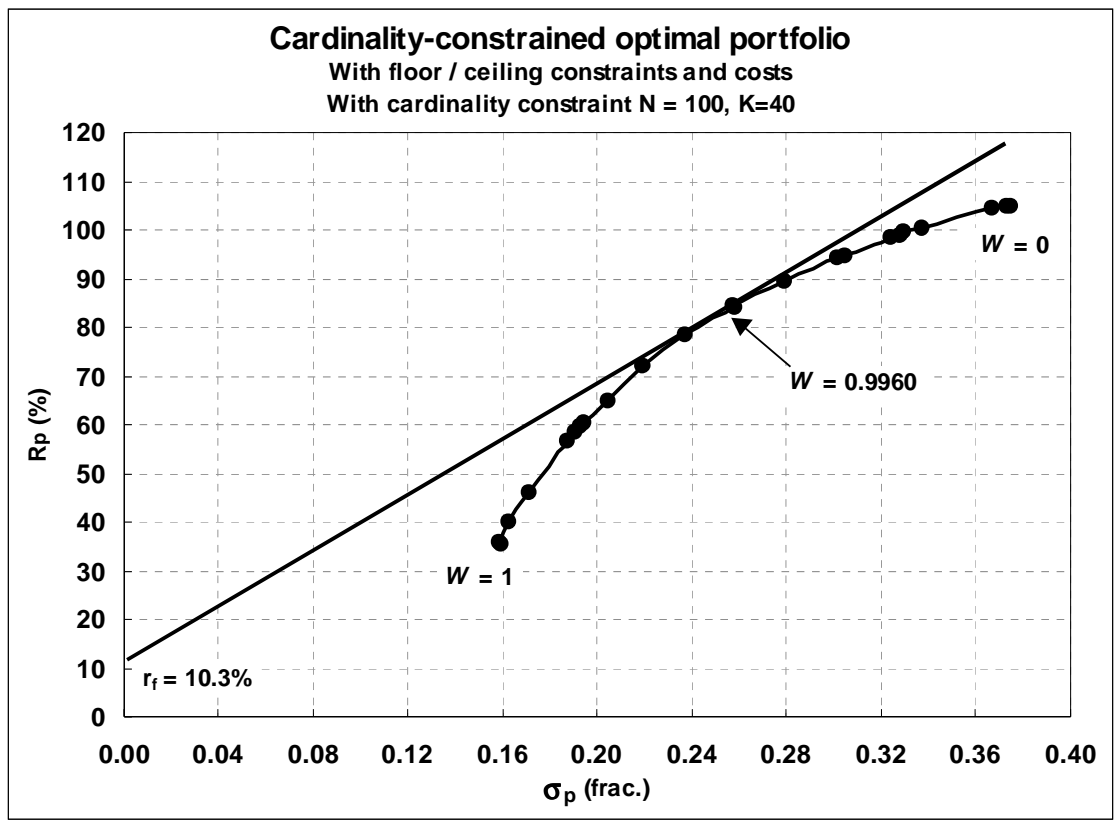

*Figure 22: The optimal cardinality-constrained portfolio*

The determination of the optimal value of *w* is shown in Figure 22 on page 53. Using this value of *w* the optimal 40-stock portfolio at the time of the study was then generated from the 100-stock universe, and is presented in Table 13.

The characteristics of the optimal 40-stock portfolio are compared with those of its parent 100-stock universe in Table 14 on page 55, which also shows the effect of then changing the floor constraint from 0,005 to 0,020. It should be noted that since the higher floor constraint shifts the frontier, a different optimal value of *w* arises.

| | Optimal cardinality-constrained portfolios | | | | |
|---|---|---|---|---|---|
| | Weight | Total costs | Excess return less costs | Beta (vs Alsi) | Forecast variance |
| | $x_i$ | $c(x)$ | $R_i$ | $\beta_i$ | $\sigma_{e_i}$ |
| | (frac) | (%) | (%) | (x) | (frac) |
| **100-stock universe** | | | | | |
| Maximum | 0.150 | 1.72 | 146.5 | 1.81 | 1.17 |
| Average | 0.010 | 0.29 | 40.2 | 1.11 | 0.14 |
| Minimum | 0.000 | 0.00 | -10.4 | 0.26 | 0.04 |
| "Portfolio" | - | 0.60 | 39.9 | 1.11 | 0.31 |
| **40-stock optimal portfolio** | | | | | |
| **Parameters:** | | Floor = | 0.005 | w = | 0.9960 |
| Maximum | 0.150 | 1.72 | 146.5 | 1.61 | 0.48 |
| Average | 0.025 | 0.73 | 59.5 | 1.00 | 0.15 |
| Minimum | 0.005 | 0.59 | 17.5 | 0.26 | 0.07 |
| Portfolio | - | 1.28 | 83.5 | 0.89 | 0.25 |
| **Parameters:** | | Floor = | 0.020 | w = | 0.9980 |
| Maximum | 0.150 | 1.72 | 146.8 | 1.61 | 0.48 |
| Average | 0.025 | 0.65 | 56.5 | 0.99 | 0.15 |
| Minimum | 0.020 | 0.62 | -4.6 | 0.26 | 0.07 |
| Portfolio | - | 0.79 | 60.1 | 0.90 | 0.25 |

*Table 13: Optimal portfolio characteristics*

| | Optimal cardinality-constrained portfolio Floor = .005, N = 100, K = 40 | | | | | | |
|---|---|---|---|---|---|---|---|
| Asset No. | Asset name | Weight | Number of assets | Forecast return | Total costs | Excess return less costs | Beta (vs Alsi) | Forecast variance |
| i | | $x_i$ (frac) | $z_i$ | $r_i$ (%) | $c(x)$ (%) | $R_i$ (%) | $\beta_i$ (x) | $\sigma_{e_i}$ (frac) |
| 1 | Iscor | 0.150 | 1 | 84.5 | 1.72 | 69.8 | 0.40 | 0.153 |
| 2 | Comparex | 0.150 | 1 | 117.8 | 1.72 | 103.1 | 1.02 | 0.150 |
| 3 | Northam | 0.150 | 1 | 62.2 | 1.72 | 47.5 | 0.26 | 0.193 |
| 4 | Educor | 0.117 | 1 | 112.4 | 1.30 | 98.0 | 1.10 | 0.142 |
| 5 | Tourvest | 0.108 | 1 | 99.8 | 1.21 | 85.6 | 0.94 | 0.144 |
| 6 | Outsors | 0.076 | 1 | 160.4 | 0.91 | 146.5 | 1.61 | 0.482 |
| 7 | Spescom | 0.051 | 1 | 138.5 | 0.74 | 124.8 | 1.53 | 0.155 |
| 8 | CCH | 0.038 | 1 | 124.6 | 0.68 | 110.9 | 1.30 | 0.292 |
| 9 | Ixchange | 0.005 | 1 | 115.6 | 0.59 | 102.0 | 1.28 | 0.372 |
| 10 | Datatec | 0.005 | 1 | 118.6 | 0.59 | 105.0 | 1.32 | 0.186 |
| 11 | Implats | 0.005 | 1 | 57.2 | 0.59 | 43.6 | 0.62 | 0.130 |
| 12 | Tiger Wheels | 0.005 | 1 | 83.3 | 0.59 | 69.7 | 0.91 | 0.118 |
| 13 | AECI | 0.005 | 1 | 62.9 | 0.59 | 49.3 | 0.83 | 0.172 |
| 14 | Unitrans | 0.005 | 1 | 61.6 | 0.59 | 48.0 | 0.78 | 0.119 |
| 15 | Amplats | 0.005 | 1 | 37.3 | 0.59 | 23.7 | 0.48 | 0.095 |
| 16 | Kersaf | 0.005 | 1 | 63.5 | 0.59 | 49.9 | 0.76 | 0.115 |
| 17 | Rembrandt | 0.005 | 1 | 67.7 | 0.59 | 54.1 | 0.96 | 0.078 |
| 18 | Afharv | 0.005 | 1 | 80.9 | 0.59 | 67.3 | 1.21 | 0.207 |
| 19 | Peregrine | 0.005 | 1 | 90.7 | 0.59 | 77.1 | 1.30 | 0.152 |
| 20 | Woolworths | 0.005 | 1 | 51.4 | 0.59 | 37.8 | 0.72 | 0.138 |
| 21 | Softline | 0.005 | 1 | 71.4 | 0.59 | 57.8 | 1.16 | 0.180 |
| 22 | Altech | 0.005 | 1 | 58.4 | 0.59 | 44.9 | 0.90 | 0.149 |
| 23 | RAHold | 0.005 | 1 | 67.6 | 0.59 | 54.0 | 1.04 | 0.154 |
| 24 | Avis | 0.005 | 1 | 59.5 | 0.59 | 45.9 | 0.87 | 0.095 |
| 25 | Malbak | 0.005 | 1 | 57.4 | 0.59 | 43.8 | 0.84 | 0.115 |
| 26 | Illovo | 0.005 | 1 | 31.1 | 0.59 | 17.5 | 0.64 | 0.100 |
| 27 | Leisurenet | 0.005 | 1 | 65.2 | 0.59 | 51.6 | 1.13 | 0.147 |
| 28 | Didata | 0.005 | 1 | 51.7 | 0.59 | 38.1 | 0.90 | 0.118 |
| 29 | OTK | 0.005 | 1 | 52.7 | 0.59 | 39.1 | 0.93 | 0.095 |
| 30 | Sasol | 0.005 | 1 | 55.0 | 0.59 | 41.4 | 1.21 | 0.115 |
| 31 | De Beers | 0.005 | 1 | 60.9 | 0.59 | 47.3 | 0.97 | 0.092 |
| 32 | Santam | 0.005 | 1 | 56.5 | 0.59 | 42.9 | 1.03 | 0.082 |
| 33 | Billiton | 0.005 | 1 | 63.2 | 0.59 | 49.6 | 1.15 | 0.104 |
| 34 | Cadschweppes | 0.005 | 1 | 33.8 | 0.59 | 20.2 | 0.57 | 0.071 |
| 35 | ABI | 0.005 | 1 | 48.3 | 0.59 | 34.7 | 0.82 | 0.102 |
| 36 | Netcare | 0.005 | 1 | 57.3 | 0.59 | 43.7 | 1.49 | 0.167 |
| 37 | Liberty | 0.005 | 1 | 60.8 | 0.59 | 47.2 | 1.05 | 0.076 |
| 38 | BOE | 0.005 | 1 | 66.3 | 0.59 | 52.7 | 1.24 | 0.110 |
| 39 | Johnnic | 0.005 | 1 | 46.8 | 0.59 | 33.2 | 1.32 | 0.132 |
| 40 | Pepkor | 0.005 | 1 | 75.9 | 0.59 | 62.3 | 1.31 | 0.163 |
| Portfolio sum/average | | 1.000 | 40 | 98 | 1.28 | **83.5** | 0.89 | **0.25** |

*Table 14: Cardinality-constrained optimal portfolio*

The optimal 40-stock portfolio has three stocks at the ceiling of 15%, five ranging in weight from 3,8% to 11,7% and the remainder at the floor of 0,5%. In contrast the portfolio with the higher floor has one stock at the ceiling, five ranging in weight from 2,1% to 5,6% and the rest at the floor of 2%. The restrictive effect of narrowing the allowable range of asset weightings is readily apparent.

The 40-stock portfolio selects stocks with above-average after-cost excess returns (averaging 59,5% compared with the universe's average of 40,2%) and below-average risk (with an average beta of 1,00 compared with 1,11). Shares with low returns and high betas and variances are excluded from the portfolio. The higher weightings in the higher-return shares results in higher costs than the notional equally-weighted "portfolio" universe, since these costs increase exponentially. This average cost of 0,73% versus 0,29% results in the total cost of the portfolio being over double that of the "portfolio" universe, at 1,28%.

For the 40-stock portfolio with the higher floor constraint, the flatter weighting distribution results in a lower average portfolio cost of 0,79% against 1,28%.

*However, the portfolio's after-cost excess return is lower by over 23% for the same level of risk.*

*Another way of looking at this severe negative impact is fairly straightforward - applying a 2% floor constraint to a 40-stock portfolio has determined 80% of the total allocation, thus leaving only a fifth of the portfolio available to be 'optimised'. Every 2% applied as a floor is 2% less that can be given to the highest-returning stocks in the portfolio.*

Clearly, restricting a portfolio's allowed weighting range can have a major detrimental effect on performance and should not be undertaken lightly or automatically. In particular the floor constraint's conventional level of 1%-2% should be re-examined relative to its associated administration and monitoring costs with a view to lowering it if at all possible.

Understanding how the optimisation proceeds is crucial to understanding the damage done by floor constraints and also counters the common knee-jerk reaction of asset managers to small weightings.

The optimal portfolio usually consists of relatively few assets with high weights which are at, or close to, the ceiling constraint, a larger but still relatively small number of 'medium' weightings and a long tail consisting of many stocks at, or close to, the floor constraint. In terms of the number of assets, this tail can be around 70%-80% of the portfolio. What happens is that the highest weightings are usually allocated to assets with high forecast (after-cost, excess) returns. However, these assets normally also have above-average risk, which raises the portfolio's risk level. This risk is then diversified away by the large number of assets with very low weightings.

- *Asset managers often query low asset weightings, on the basis that their impact on the portfolio's returns will be negligible. This is quite correct. However, their effectiveness lies not in raising returns, but in reducing risk through diversification.*

The route through which floor constraints damage portfolio performance now becomes readily apparent. By raising the floor, there will be fewer of these 'tail-end' stocks. The diversification and risk reduction effected by this portion of the portfolio is thus reduced. Therefore the portfolio's overall risk level can only be reduced by reducing the high weightings in the high-risk, high-return assets. This then reduces the portfolio's return for the same level of risk. Alternatively, the portfolio would have been riskier if its return had been left unchanged.

This suggests there may be a tendency for institutional investors to overconstrain portfolios with a plethora of judgemental policy guidelines which include legal requirements, "prudential" rules and market factors such as tradability, as well as deviations from benchmark structures and even competitors' portfolios.

Often, after compliance with all these constraints, the opportunity remaining for any optimisation has effectively been crowded out.

# 5. Conclusion and recommendations

## 5.1 Conclusions

In this thesis a general model for the optimisation of realistic portfolios has been presented.

It must be noted that the technique developed is applicable to portfolios consisting of mixtures of any type of asset, as long as return and risk forecasts are available.

The research has shown that realistically large portfolios which, in addition to floor and ceiling constraints, contain
- nonlinear transaction costs, including a substantial illiquidity premium; and
- cardinality constraints

can be optimised effectively and in reasonable times using heuristic algorithms.

These two real-life elements are generally not found in commercial portfolio optimisation packages.

Of the heuristics tested, the performance of genetic algorithms was orders of magnitude better than that for tabu/scatter search for this application and problem size.

The GA heuristic applied to portfolio optimisation is effective and robust with respect to:
- quality of solutions;
- speed of convergence;
- versatility in not relying on any assumed or restrictive properties of the model;
- the easy addition of new constraints; and
- the easy modification of the objective function (e.g. the incorporation of higher moments than the variance or the use of alternative risk measures such as Sortino/downside risk).

The flexibility of the model is markedly greater than for some commercial portfolio optimisation packages, although in its current form it does not offer the same amount of integration and ease of use, particularly in data generation.

The usual negative aspect of metaheuristic methods, the need for tailoring, customising and fine-tuning the algorithm, was not an issue. While this would no doubt have improved the performance of the model to some extent, it was not found necessary and not undertaken in this application. The performance of the GA was resilient with regard to parameter settings.

Some of the insights gained from the research were:
- Both floor and ceiling constraints have a substantial negative impact on portfolio performance and should be examined critically relative to their associated administration and monitoring costs;
- The optimal portfolio with cardinality constraints often contains a large number of stocks with very low weightings.
- Asset managers' knee-jerk objection to low weightings on the basis that they do not benefit returns materially is misplaced, since their function is not to raise returns but to reduce risk. Unnecessarily high floor constraints interfere with this function and damage portfolio performance severely.
- Nonlinear transaction costs which are comparable to forecast returns in magnitude will tend to diversify portfolios materially; the effect of these costs on portfolio risk is ambiguous, depending on the degree of diversification required for cost reduction;
- The number of assets in a portfolio invariably increases as a result of constraints, costs and their combination.
- The optimal portfolios generated for the cardinality-constrained case will always be conservative relative to the true efficient frontier.
- The implementation of cardinality constraints is essential for finding the best-performing portfolio. The ability of the heuristic method to deal with cardinality constraints is one of its most powerful features.

## 5.2 Recommendations

Further work is suggested in the following areas.

Clearly, individual stocks will suffer different illiquidity premiums. The model can be refined by providing individual cost curves for each stock. Implementation is easy, but estimating the illiquidity premium is difficult.

Similarly, individual floor and ceiling constraints can be applied to each asset class, and other relevant constraints would relate to an asset's market capitalisation, tradability, or both.

Style, class or sector constraints can be added to the model. These constraints limit the proportion of the portfolio that can be invested in shares which fall into a style definition (e.g. value/growth, cyclical/defensive, smallcap, liquid, rand-hedge etc.) or a market sector.

Different cardinality-constrained efficient frontiers will be generated for different values of $K$. Clearly, as $K$ decreases (relative to the total number of stocks in the universe, $N$) the portfolio's potential performance (albeit with higher risk) increases and the frontier will move further away (upwards) from the cardinality-unconstrained efficient frontier. The magnitude of the sensitivity of this movement to different values of the ratio ($K/N$) is worth investigating.

The input forecasts for return and risk are point forecasts, making the model deterministic. A stochastic approach could be taken by attaching distributions to the input forecasts, resulting in an objective function which is also a distribution. While it is usually the mean which will be optimised, its variance can also be monitored.

- oOo -

# 9. Bibliography

## 9.1 Portfolio theory

## 9.2 General heuristic methods

## 9.3 Genetic algorithms

## 9.4 Tabu search

# 8. Appendices

# Appendix I: Genetic algorithms

(This appendix is sourced largely from the paper by Beasley, Bull and Martin, which is referenced in the bibliography.)

## 1. The method

### 1.1 Overview

The execution of the genetic algorithm is a two-stage process. It starts with the current population. Selection is applied to the current population to create an intermediate population. Then *recombination* and *mutation* are applied to the intermediate population to create the next population. The process of going from the current population to the next population constitutes one *generation* in the execution of a genetic algorithm.

The *evaluation function*, or *objective function*, provides a measure of performance with respect to a particular set of parameters. The *fitness function* transforms that measure of performance into an allocation of reproductive opportunities. The evaluation of a string representing a set of parameters is independent of the evaluation of any other string. The fitness of that string, however, is always defined with respect to other members of the current population. In a genetic algorithm, fitness is defined by $f_i/f_A$ where $f_i$ is the evaluation associated with string $i$ and $f_A$ is the average evaluation of all the strings in the population. Fitness can also be assigned based on a string's *rank* in the population or by sampling methods, such as *tournament selection*.

The standard GA can be represented as shown in Figure 23 on page 66.

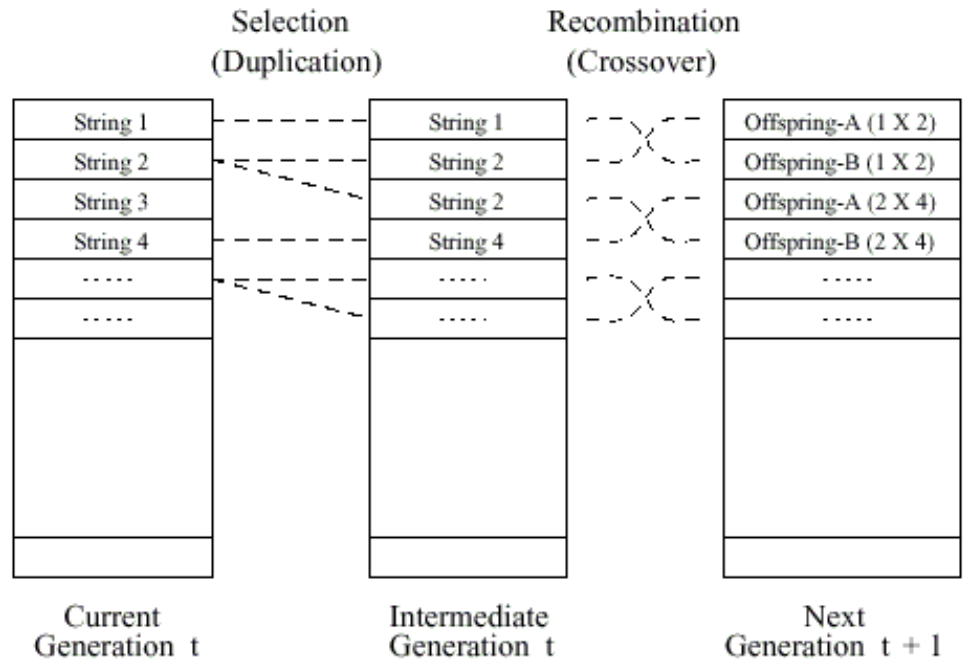

*Figure 23: GA process*

In the first generation the current population is also the initial population. After calculating $f_i/f_A$ for all the strings in the current population, selection is carried out. The probability that strings in the current population are copied (i.e. duplicated) and placed in the intermediate generation is in proportion to their fitness.

Individuals are chosen using *"stochastic sampling with replacement"* to fill the intermediate population. A selection process that will more closely match the expected fitness values *is "remainder stochastic sampling"*. For each string $i$ where $f_i/f_A$ is greater than 1,0, the integer portion of this number indicates how many copies of that string are placed directly in the intermediate population. All strings (including those with $f_i/f_A$ less than 1,0) then place additional copies in the intermediate population with a probability corresponding to the fractional portion of $f_i/f_A$. For example, a string with $f_i/f_A = 1,36$ places 1 copy in the intermediate population, and then receives a 0,36 chance of placing a second copy. A string with a fitness of $f_i/f_A = 0,54$ has a 0,54 chance of placing one string in the intermediate population. Remainder stochastic sampling is most efficiently implemented using a method known as *stochastic universal sampling*. Assume that the population is laid out in random order as in a pie graph, where each individual is assigned space on the pie graph in proportion to fitness. An outer roulette wheel is placed around the pie with *N* equally-spaced pointers. A single spin of the roulette wheel will now simultaneously pick all *N* members of the intermediate population.

After selection has been carried out the construction of the intermediate population is complete and *recombination* can occur. This can be viewed as creating the next population from the intermediate population. *Crossover* is applied to randomly paired strings with a probability denoted $p_c$. (The population should already be sufficiently shuffled by the random selection process.) Pick a pair of strings. With probability $p_c$ "*recombine*" these strings to form two new strings that are inserted into the next population.

Consider the following binary string: 1101001100101101. The string could represent a possible solution to some parameter optimisation problem. New sample points in the space are generated by recombining two parent strings. Consider this string 1101001100101101 and another binary string, *yxyyxyxxyyyxyxxy*, in which the values 0 and 1 are denoted by x and y. Using a single randomly-chosen recombination point, one-point crossover occurs as follows:

$$11010 \setminus / 01100101101$$
$$yxyyx / \setminus yxxyyyxyxxy$$

Swapping the fragments between the two parents produces the following offspring:

$$11010yxxyyyxyxxy \quad \text{and} \quad yxyyx01100101101$$

After recombination, we can apply a *mutation operator*. For each bit in the population, mutate with some low probability $p_m$. Typically the mutation rate is applied with 0,1%-1,0% probability. After the process of selection, recombination and mutation is complete, the next population can be evaluated. The process of valuation, selection, recombination and mutation forms one *generation* in the execution of a genetic algorithm.

## 1.2 Coding

Before a GA can be run, a suitable *coding* (or representation) for the problem must be devised. We also require a fitness function, which assigns a figure of merit to each coded solution. During the run, parents must be selected for reproduction, and recombined to generate offspring.

It is assumed that a potential solution to a problem may be represented as a set of parameters (for example, the parameters that optimise a neural network). These parameters (known as *genes*) are joined together to form a string of values (often referred to as a *chromosome*. For example, if the problem is to maximise a function of three variables, $F(x; y; z)$, we might represent each variable by a 10-bit binary number (suitably scaled). Our chromosome would therefore contain three genes, and consist of 30 binary digits. The set of parameters represented by a particular chromosome is referred to as a *genotype*. The genotype contains the information required to construct an organism which is referred to as the *phenotype*. For example, in a bridge design task, the set of parameters specifying a particular design is the genotype, while the finished construction is the phenotype.

The fitness of an individual depends on the performance of the phenotype. This can be inferred from the genotype, i.e. it can be computed from the chromosome, using the fitness function. Assuming the interaction between parameters is nonlinear, the size of the search space is related to the number of bits used in the problem encoding. *For a bit string encoding of length L; the size of the search space is $2^L$ and forms a hypercube*. The genetic algorithm samples the corners of this *L*-dimensional hypercube. Generally, most test functions are at least 30 bits in length; anything much smaller represents a space which can be *enumerated.* Obviously, the expression $2^L$ grows exponentially. *As long as the number of "good solutions" to a problem is sparse with respect to the size of the search space, then random search or search by enumeration of a large search space is not a practical form of problem solving*. On the other hand, any search other than random search imposes some bias in terms of how it looks for better solutions and where it looks in the search space. A genetic algorithm belongs to the class of methods known as *"weak methods"* because it makes relatively few assumptions about the problem that is being solved. Genetic algorithms are often described as a *global search method* that does not use *gradient information*. Thus, *nondifferentiable functions* as well as functions with *multiple local optima* represent classes of problems to which genetic algorithms might be applied. Genetic algorithms, as a weak method, are robust but very general.

## 1.3 Fitness function

A fitness function must be devised for each problem to be solved. Given a particular chromosome, the fitness function returns a single numerical "fitness" or "figure of merit" which is supposed to be proportional to the "utility" or "ability" of the individual which that chromosome represents. For many problems, particularly function optimisation, the fitness function should simply measure the value of the function.

## 1.4 Reproduction

Good individuals will probably be selected several times in a generation, poor ones may not be at all. Having selected two parents, their chromosomes are recombined, typically using the mechanisms of *crossover* and *mutation*. The previous crossover example is known as *single point crossover*. Crossover is not usually applied to all pairs of individuals selected for mating. A random choice is made, where the likelihood of crossover being applied is typically between 0,6 and 1,0. If crossover is not applied, offspring are produced simply by duplicating the parents. This gives each individual a chance of passing on its genes without the disruption of crossover.

*Mutation* is applied to each child individually after crossover. It randomly alters each gene with a small probability. The following diagram shows the fifth gene of a chromosome being mutated:

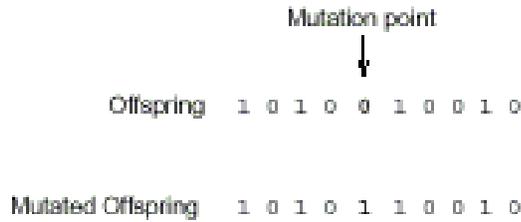

The traditional view is that crossover is the more important of the two techniques for rapidly *exploring* a search space. Mutation provides a small amount of *random search*, and helps ensure that no point in the search has a zero probability of being examined.

An example of two individuals reproducing to give two offspring is shown in Figure 24.

| Individual | Value | Fitness | Chromosome |
|---|---|---|---|
| Parent 1 | 0.08 | 0.05 | 00 01010010 |
| Parent 2 | 0.73 | 0.000002 | 10 11101011 |
| Offspring 1 | 0.23 | 0.47 | 00 11101011 |
| Offspring 2 | 0.58 | 0.00007 | 10 01010010 |

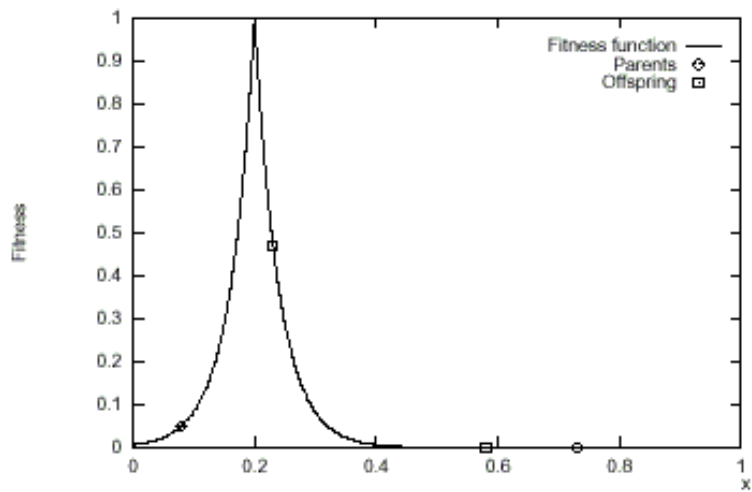

*Figure 24: Illustration of crossover*

The fitness function is an exponential function of one variable, with a maximum at $x = 0,2$. It is coded as a 10-bit binary number. This illustrates how it is possible for crossover to recombine parts of the chromosomes of two individuals and give rise to offspring of higher fitness. (Crossover can also produce offspring of low fitness, but these will not be likely to be selected for reproduction in the next generation.)

## 1.5 Convergence

The fitness of the best and the average individual in each generation increases towards a global optimum. *Convergence* is the progression towards increasing uniformity. A gene is said to have converged when 95% of the population share the same value. The population is said to have converged when all of the genes have converged. As the population converges, the average fitness will approach that of the best individual.

A GA will always be subject to stochastic errors. One such problem is that of *genetic drift*. Even in the absence of any selection pressure (i.e. a constant fitness function), members of the population will still converge to some point in the solution space. This happens simply because of the accumulation of stochastic errors. If, by chance, a gene becomes *predominant* in the population, then it is just as likely to become more predominant in the next generation as it is to become less predominant. If an increase in predominance is sustained over several successive generations, and the population is finite, then a gene can spread to all members of the population. Once a gene has converged in this way, it is fixed; crossover cannot introduce new gene values. This produces a ratchet effect, so that as generations go by, each gene eventually becomes fixed. The rate of genetic drift therefore provides a lower bound on the rate at which a GA can converge towards the correct solution. That is, if the GA is to exploit gradient information in the fitness function, the fitness function must provide a slope sufficiently large to counteract any genetic drift. The rate of genetic drift can be reduced by increasing the mutation rate. However, if the mutation rate is too high, the search becomes effectively random, so once again gradient information in the fitness function is not exploited.

## 2.3 Strengths and weaknesses

### 2.3.1 Strengths

The power of GAs comes from the fact that the technique is robust and can deal successfully with a wide range of difficult problems.

GAs are not guaranteed to find the global optimum solution to a problem, but they are generally good at finding "acceptably good" solutions to problems "acceptably quickly". Where specialised techniques exist for solving particular problems, they are likely to outperform GAs in both speed and accuracy of the final result.

Even where existing techniques work well, improvements have been realised by hybridising them with a GA.

The basic mechanism of a GA is so robust that, within fairly wide margins, parameter settings are not critical.

### 2.3.2 Weaknesses

A problem with GAs is that the genes from a few comparatively highly fit (but not optimal) individuals may rapidly come to dominate the population, causing it to converge on a local maximum. Once the population has converged, the ability of the GA to continue to search for better solutions is effectively eliminated: crossover of almost identical chromosomes produces little that is new. Only mutation remains to explore entirely new ground, and this simply performs a slow, random search.

### 2.4 Applicability

Most traditional GA research has concentrated in the area of *numerical function optimisation*. GAs have been shown to be able to outperform conventional optimisation techniques on difficult, discontinuous, multimodal, noisy functions. *These characteristics are typical of market data, so this technique is well suited to the objective of asset allocation.*

For asset allocation, combinatorial optimisation requires solutions to problems involving arrangements of discrete objects. This is quite unlike function optimisation, and different coding, recombination, and fitness function techniques are required.

There are many applications of GAs to learning systems, the usual paradigm being that of a classifier system. The GA tries to evolve (i.e. learn) a set of "if : : : then" rules to deal with some particular situation. This has been applied to *economic modelling* and *market trading* [2], once again our area of interest.

## 2.5 Practical implementation

### 2.5.1 Fitness function

Along with the *coding scheme* used, *the fitness* function is the most crucial aspect of any GA. Ideally, the fitness function should be smooth and regular, so that chromosomes with reasonable fitness are distinguishable from chromosomes with slightly better fitness. They should not have too many local maxima, or a very isolated global maximum. It should reflect the value of the chromosome in some "real" way, but unfortunately *the "real" value of a chromosome is not always a useful quantity for guiding genetic search*. In combinatorial optimisation problems, where there are many constraints, most points in the search space often represent invalid chromosomes and hence have zero "real" value. Another approach which has been taken in this situation is to use a *penalty function,* which represents how poor the chromosome is, and *construct the fitness as (constant - penalty).* A suitable form is:

$$f_a(x) = f(x) + M^k w^T c_v(x)$$

where **w** is a vector of nonnegative weighting coefficients, the vector $c_V$ quantifes the magnitudes of any constraint violations, $M$ is the number of the current generation and $k$ is a suitable exponent. The dependence of the penalty on generation number biases the search increasingly heavily towards feasible space as it progresses.

Penalty functions which represent the *amount* by which the constraints are violated are better than those which are based simply on the *number* of constraints which are violated.

*Approximate function evaluation* is a technique which can sometimes be used if the fitness function is excessively slow or complex to evaluate. A GA should be robust enough to be able to converge in the face of the noise represented by the approximation. Approximate fitness techniques have to be used in cases where the fitness function is stochastic.

### 2.5.2 Fitness Range Problems

**Premature convergence**
*The initial population may be generated randomly, or using some heuristic method.* At the start of a run, the values for each gene for different members of the population are distributed randomly. Consequently, there is a wide spread of individual fitnesses. As the run progresses, particular values for each gene begin to predominate. As the population converges, so the range of fitnesses in the population reduces. *This variation in fitness range throughout a run often leads to the problems of premature convergence and slow finishing.*

Holland's [24] *schema theorem* says that one should allocate reproductive opportunities to individuals in proportion to their relative fitness. But then premature convergence occurs because the population is not infinite. To make GAs work effectively on finite populations, the way individuals are selected for reproduction must be modified. One needs to control the number of reproductive opportunities each individual gets so that it is neither too large nor too small. The effect is to *compress the range of fitnesses*, and prevent any "super-fit" individuals from suddenly taking over.

**Slow finishing**
This is the converse problem to premature convergence. After many generations, the population will have largely converged, but may still not have located the global maximum precisely. The average fitness will be high, and there may be little difference between the best and the average individuals. Consequently *there is an insufficient gradient in the fitness function to push the GA towards the maximum.*

The same techniques which are used to combat premature convergence also combat slow finishing. They do this by *expanding the effective range of fitnesses* in the population. As with premature convergence, fitness scaling can be prone to overcompression due to just one "super poor" individual.

### 2.5.3 Parent selection techniques

*Parent selection* is the task of allocating reproductive opportunities to each individual. In principle, individuals from the population are copied to a *"mating pool"*, with highly-fit individuals being more likely to receive more than one copy, and unfit individuals being more likely to receive no copies. Under a strict generational replacement, the size of the mating pool is equal to the size of the population. After this, pairs of individuals are taken out of the mating pool at random, and mated. This is repeated until the mating pool is exhausted. The behaviour of the GA very much depends on how individuals are chosen to go into the mating pool. Ways of doing this can be divided into two methods:

1) **Explicit fitness remapping**
To keep the mating pool the same size as the original population, the average of the number of reproductive trials allocated per individual must be one. If each individual's fitness is remapped by dividing it by the average fitness of the population, this effect is achieved. This remapping scheme allocates reproductive trials in proportion to raw fitness, according to Holland's theory. The remapped fitness of each individual will, in general, not be an integer. Since only an integral number of copies of each individual can be placed in the mating pool, we have to convert the number to an integer in a way that does not introduce bias. A better method than *stochastic remainder sampling without replacement* is *stochastic universal sampling,* which is elegantly simple and theoretically perfect. It is important not to confuse the sampling method with the parent selection method. Different parent selection methods may have advantages in different applications. But a good sampling method is always good, for all selection methods, in all applications.

*Fitness scaling* is a commonly employed method of remapping. The maximum number of reproductive trials allocated to an individual is set to a certain value, typically 2,0. This is achieved by subtracting a suitable value from the raw fitness score, then dividing by the average of the adjusted fitness values. Subtracting a fixed amount increases the ratio of maximum fitness to average fitness. Care must be taken to prevent negative fitness values being generated. However, the presence of *just one super-fit* individual (with a fitness ten times greater than any other, for example), can lead to *overcompression*. If the fitness scale is compressed so that the ratio of maximum to average is 2:1, then the rest of the population will have fitnesses clustered closely about 1. Although premature convergence has been prevented, it has been at the expense of effectively flattening out the fitness function. As mentioned previously, if the fitness function is too flat, *genetic drift* will become a problem, so overcompression may lead not just to slower performance, but also to drift away from the maximum.

*Fitness windowing* is the same as fitness scaling, except the amount subtracted is the minimum fitness observed during the previous $n$ generations, where $n$ is typically 10. With this scheme the selection pressure (i.e. the ratio of maximum to average trials allocated) varies during a run, and also from problem to problem. The presence of a super-unfit individual will cause underexpansion, while super-fit individuals may still cause premature convergence, since they do not influence the degree of scaling applied. The problem with both fitness scaling and fitness windowing is that the degree of compression is dictated by a single, extreme individual, either the fittest or the worst. Performance will suffer if the extreme individual is exceptionally extreme.

*Fitness ranking* is another commonly-employed method, which overcomes the reliance on an extreme individual. Individuals are sorted in order of raw fitness, and then reproductive fitness values are assigned according to rank. This may be done linearly or exponentially. This gives a similar result to fitness scaling, in that the ratio of the maximum to average fitness is normalised to a particular value. However it also ensures that the remapped fitnesses of intermediate individuals are regularly spread out. Because of this, the effect of one or two extreme individuals will be negligible, irrespective of how much greater or less their fitness is than the rest of the population. The number of reproductive trials allocated to, say, the fifth-best individual will always be the same, whatever the raw fitness values of those above (or below). The effect is that overcompression ceases to be a problem. Several experiments have shown ranking to be superior to fitness scaling.

### 2. Implicit fitness remapping

Implicit fitness remapping methods fill the mating pool without passing through the intermediate stage of remapping the fitness.

In *binary tournament selection*, pairs of individuals are picked at random from the population. Whichever has the higher fitness is copied into a mating pool (and then both are replaced in the original population). This is repeated until the mating pool is full. Larger tournaments may also be used, where the best of *n* randomly-chosen individuals is copied into the mating pool. Using larger tournaments has the effect of increasing the *selection pressure*, since below-average individuals are less likely to win a tournament and vice-versa.

A further generalisation is *probabilistic binary tournament selection*. In this, the better individual wins the tournament with probability *p*, where $0,5 < p < 1$. Using lower values of p has the effect of decreasing the selection pressure, since below-average individuals are comparatively more likely to win a tournament and vice-versa. By adjusting tournament size or win probability, the selection pressure can be made arbitrarily large or small.

## 2.5.4 Other crossovers

**Two-point crossover**
The problem with adding additional crossover points is that *building blocks are more likely to be disrupted*. However, an advantage of having more crossover points is that *the problem space may be searched more thoroughly*. In *two-point crossover*, (and multi-point crossover in general), rather than linear strings, chromosomes are regarded as *loops* formed by joining the ends together. To exchange a segment from one loop with that from another loop requires the selection of two cut points, as shown in Figure 25:

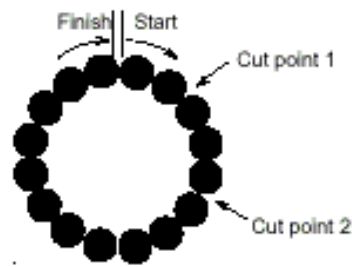

*Figure 25: Two-point crossover*

Here, one-point crossover can be seen as two-point crossover with one of the cut points fixed at the start of the string. Hence two-point crossover performs the same task as one-point crossover (i.e. exchanging a single segment), but is more general. A chromosome considered as a loop can contain more building blocks since they are able to "wrap around" at the end of the string. two-point crossover is generally better than one-point crossover.

**Uniform crossover**
*Uniform crossover* is radically different to one-point crossover. Each gene in the offspring is created by copying the corresponding gene from one or the other parent, chosen according to a randomly generated *crossover mask*. Where there is a 1 in the crossover mask, the gene is copied from the first parent, and where there is a 0 in the mask, the gene is copied from the second parent, as follows:

```
Crossover Mask   1 0 0 1 0 1 1 1 0 0
      Parent 1   1 0 1 0 0 0 1 1 1 0
                 ↓     ↓   ↓ ↓ ↓
   Offspring 1   1 1 0 0 0 0 1 1 1 1
                   ↑ ↑   ↑       ↑ ↑
      Parent 2   0 1 0 1 0 1 0 0 1 1
```

The process is repeated with the parents exchanged to produce the second offspring. A new crossover mask is randomly generated for each pair of parents. Offspring therefore contain a mixture of genes from each parent. The number of effective crossing points is not fixed, but will average $L/2$ (where $L$ is the chromosome length).

Uniform crossover appears to be more robust. Where two chromosomes are similar, the segments exchanged by two-point crossover are likely to be identical, leading to offspring which are identical to their parents. This is less likely to happen with uniform crossover.

### 2.5.5 Inversion and Reordering

The order of genes on a chromosome is critical for the method to work effectively. Techniques for *reordering* the positions of genes in the chromosome during a run have been suggested. One such technique, *inversion*, works by reversing the order of genes between two randomly-chosen positions within the chromosome. Reordering does nothing to lower epistasis (see Section 2.5.6), but greatly expands the search space. Not only is the GA trying to find good sets of *gene values*, it is simultaneously trying to discover good *gene orderings* too.

### 2.5.6 Epistasis

*Epistasis* is the interaction between different genes in a chromosome. It is the extent to which the "*expression*" (i.e. contribution to fitness) of one gene depends on the values of other genes. The degree of interaction will be different for each gene in a chromosome. If a small change is made to one gene we expect a resultant change in chromosome fitness. This resultant change may vary according to the values of other genes.

### 2.5.7 Deception

One of the fundamental principles of GAs is that chromosomes which include schemata which are contained in the global optimum will increase in frequency (this is especially true of short, low-order schemata, known as building blocks). Eventually, via the process of crossover, these optimal schemata will come together, and the globally optimum chromosome will be constructed. But if schemata which are not contained in the global optimum increase in frequency more rapidly than those which are, the GA will be misled, away from the global optimum, instead of towards it. This is known as *deception*. Deception is a special case of epistasis and epistasis is necessary (but not sufficient) for deception. If epistasis is very high, the GA will not be effective. If it is very low, the GA will be outperformed by simpler techniques, such as hillclimbing.

### 2.5.8 Mutation and Naïve Evolution

Mutation is traditionally seen as a "background" operator, responsible for re-introducing *alleles* or inadvertently lost gene values, preventing genetic drift and providing a small element of random search in the vicinity of the population when it has largely converged. It is generally held that crossover is the main force leading to a thorough search of the problem space. *"Naïve evolution"* (just selection and mutation) performs a hillclimb-like search which can be powerful without crossover. However, mutation generally finds better solutions than a crossover-only regime. *Mutation becomes more productive, and crossover less productive, as the population converges.* Despite its generally low probability of use, mutation is a very important operator. Its optimum probability is much more critical than that for crossover. Mutation becomes more productive, and crossover less productive, as the population converges.

### 2.5.9 Niches and speciation

*Speciation* is the process whereby a single species differentiates into two (or more) different species occupying different niches. In a GA, niches are analogous to maxima in the fitness function. Sometimes we have a fitness function which is known to be multimodal, and we may want to locate all the peaks. Unfortunately a traditional GA will not do this; the whole population will eventually converge on a single peak. This is due to *genetic drift*. The two basic techniques to solve this problem are to *maintain diversity*, or to *share the payoff* associated with a niche.

In *preselection*, offspring replace the parent only if the offspring's fitness exceeds that of the inferior parent. There is fierce competition between parents and children, so the payoff is not so much shared as fought over, and the winner takes all. This method helps to maintain diversity (since strings tend to replace others which are similar to themselves) and this helps prevent convergence on a single maximum.

In a *crowding* scheme, offspring are compared with a few (typically two or three) randomly-chosen individuals from the population. The offspring replaces the most similar one found. This again aids diversity and indirectly encourages speciation.

### 2.5.10 Restricted Mating

The purpose of *restricted mating* is to encourage speciation, and reduce the production of *lethals*. A lethal is a child of parents from two different niches. Although each parent may be highly fit, the combination of their chromosomes may be highly unfit if it falls in the valley between the two maxima The general philosophy of restricted mating makes the assumption that if two similar parents (i.e. from the same niche) are mated, then the offspring will be similar. However, this will very much depend on the coding scheme and low epistasis. Under conventional crossover and mutation operators, two parents with similar genotypes will always produce offspring with similar genotypes. However, in a highly epistatic chromosome, there is no guarantee that these offspring will not be of low fitness, i.e. lethals.

The total reward available in any niche is fixed, and is distributed using a bucket-brigade mechanism. In *sharing*, several individuals which occupy the same niche are made to share the fitness payoff among them. Once a niche has reached its "*carrying capacity*", it no longer appears rewarding in comparison with other, unfilled niches.

### 2.5.11 Diploidy and Dominance

In the higher life-forms, chromosomes contain two sets of genes, rather than just one. This is *diploidy*. (A *haploid* chromosome contains only one set of genes.) Virtually all work on GAs concentrates on haploid chromosomes. This is primarily for simplicity, although use of diploid chromosomes might have benefits. Diploid chromosomes lend advantages to individuals where the environment may change over a period of time. Having two genes allows two different "solutions" to be remembered, and passed on to offspring. One of these will be *dominant* (that is, it will be expressed in the phenotype), while the other will be *recessive*. If environmental conditions change, the dominance can shift, so that the other gene is dominant. This shift can take place much more quickly than would be possible if evolutionary mechanisms had to alter the gene. This mechanism is ideal if the environment regularly switches between two states.

## 2.6 Summary

The major advantage of genetic algorithms is their *flexibility* and *robustness* as a global search method. They are "weak methods" which do not use gradient information and make relatively few assumptions about the problem being solved. They can deal with highly nonlinear problems and non-differentiable functions as well as functions with multiple local optima. They are also readily amenable to parallel implementation, which renders them usable in real-time.

The primary drawback of genetic algorithms results from their flexibility. The designer has to come up with encoding schemes that allow the GA to take advantage of the underlying building blocks. One has to make sure the evaluation function assigns meaningful fitness measures to the GA. It is not always clear how the evaluation function can be formulated for the GA to produce an optimal solution. GAs are also computationally intensive and convergence is sometimes a problem.

GAs are highly effective in modelling asset allocation problems, because the driving variables are highly nonlinear, noisy, chaotic and changing all the time. (They are, however, well-established and relatively few.)

# Appendix II: Tabu and scatter search

(This appendix is sourced largely from the book by Michalewicz and Fogel [25] and the paper by Glover, Kelley and Laguna [26].)

## 1. Scatter search

Parallel to the development of GAs, Glover established the principles and operational rules for *tabu search* (TS) and a related methodology known as *scatter search* [27].

Scatter search has some interesting commonalties with GA ideas, although it also has a number of quite distinct features. Several of these features have come to be incorporated into GA approaches after an intervening period of approximately a decade, while others remain largely unexplored in the GA context.

Scatter search [28] is designed to operate on a set of points, called *reference points,* that constitute good solutions obtained from previous solution efforts. The approach systematically generates *linear combinations* of the reference points to create new points, each of which is mapped into an associated feasible point. Tabu search is then superimposed to control the composition of reference points at each stage. Tabu search has its roots in the field of artificial intelligence as well as in the field of optimisation.

The heart of tabu search lies in its use of *adaptive memory*, which provides the ability to take advantage of the search history in order to guide the solution process. In its simplest manifestations, adaptive memory is exploited to prohibit the search from reinvestigating solutions that have already been evaluated. However, the use of memory in scatter search implementation is much more complex and calls upon memory functions that encourage search *diversification* and *intensification*. These memory components allow the search to escape from locally optimal solutions and in many cases find a globally optimal solution.

Similarities are immediately evident between scatter search and the original GA proposals. Both are instances of what are sometimes called *population-based* approaches. Both incorporate the idea that a key aspect of producing new elements is to generate some form of combination of existing elements. On the other hand, several contrasts between these methods may be noted. The early GA approaches were predicated on the idea of choosing parents randomly to produce offspring, and then introducing randomisation to determine which components of the parents should be combined. In contrast, the scatter search approach does not correspondingly make recourse to randomisation, in the sense of being indifferent to choices among alternatives.

However, the approach is designed to incorporate strategic probabilistic biases, taking account of evaluations and history. Scatter search focuses on generating relevant outcomes without losing the ability to produce diverse solutions, due to the way the generation process is implemented. For example, the approach includes the generation of new points that are not convex combinations of the original points. The new points may then contain information that is not contained in the original reference points.

Scatter search is an *information-driven* approach, exploiting knowledge derived from the search space, high-quality solutions found within the space, and trajectories through the space over time. The combination of these factors creates a highly effective solution process.

## 2. Tabu and scatter search

A basic tabu algorithm is shown in Figure 26 on page 85.

One way of intelligently guiding a search process is to forbid (or discourage) certain solutions from being chosen based on information that suggests these solutions may duplicate, or significantly resemble, solutions encountered in the past. In tabu search, this is often done by defining suitable attributes of moves or solutions, and imposing restrictions on a set of the attributes, depending on the search history. Two prominent ways for exploiting search history in TS are through *recency* and *frequency* memories.

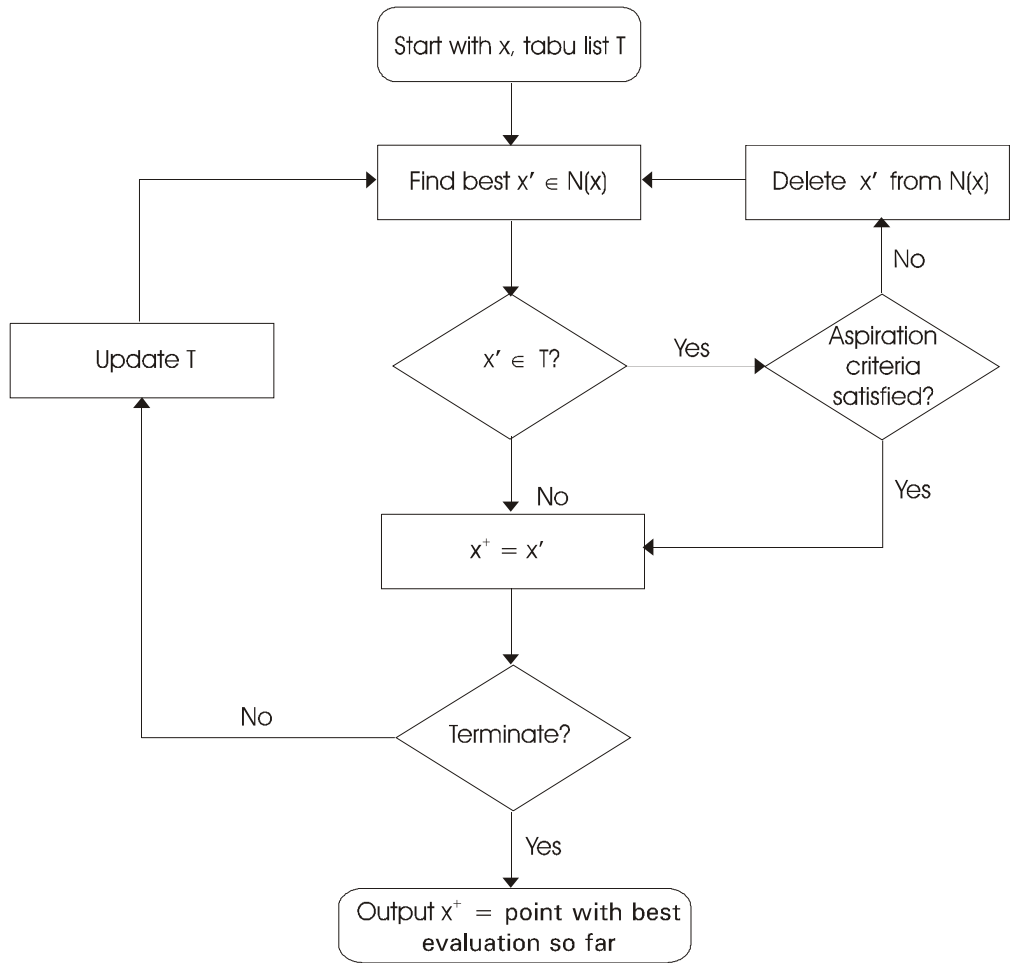

*Figure 26: TS flowsheet*

Recency memory is typically (though not invariably) a short-term memory that is managed by structures or arrays called *tabu lists*, while frequency memory more usually fulfills a long-term search function. A standard form of recency memory discourages moves that lead to solutions with attributes shared by other solutions recently visited. Frequency-based memory can be useful for diversifying the search. A standard form of frequency memory discourages moves leading to solutions whose attributes have often been shared by solutions visited during the search, or alternately encourages moves leading to solutions whose attributes have rarely been seen before. Another standard form of frequency memory is defined over subsets of elite solutions to fulfill an intensification function.

Short- and long-term components based on recency and frequency memory can be used separately or together in complementary TS search strategies. Note that this approach operates by implicitly modifying the neighborhood of the current solution.

Tabu search in general includes many enhancements to this basic scheme. The details of the short-term and long-term adaptive memories, and a recovery strategy for both intensifying and diversifying the search are discussed in Section 3.

Tabu search considers solutions from the whole neighborhood and selects a non-tabu solution as the next starting solution, regardless whether it has a better evaluation score than the current solution. But in "abnormal" circumstances, such as when an excellent but tabu solution is found in the neighborhood, this solution is accepted. This override of the tabu classification occurs when an *aspiration criterion* is met.

The *deterministic* selection procedure can also be changed into a *probabilistic* method where better solutions have an increased probability of being selected. The memory horison can be changed during the search and this memory size can also be linked to the size of the problem, e.g. remembering the last $n$ moves, where $n$ represents the size of the problem.

The use of long-term memory is usually restricted to special circumstances, such as where all non-tabu moves lead to inferior solutions, where reference to the contents of a long-term memory may be useful to decide the next search direction. A typical approach makes the most frequent moves less attractive. The evaluation score is decreased by some penalty that depends on the frequency, and the final score determines the winner.

## 3. Overview of the algorithm

We assume that a solution to the optimisation problem can be represented by a $n$-dimensional vector **x**, where $x_i$ may be a real or an integer bounded variable (for $i = 1$; ...; $n$). In addition, we assume that the objective function value $f(\mathbf{x})$ can be obtained by running a related simulation model that uses **x** as the value of its input factors. Finally, a set of linear constraints (equality or inequality) may be imposed on **x**.

The algorithm starts by generating an initial population of *reference points*. The initial population may include points suggested by the user, and it always includes the following midpoint:

$$x_i = l_i + (u_i - l_i)/2$$

where $u_i$ and $l_i$ are the upper and lower bounds on $x_i$, respectively. Additional points are generated with the goal of creating a *diverse population*. A population is considered diverse if its elements are "significantly" different from one another. We use a distance measure to determine how "close" a potential new point is from the points already in the population, in order to decide whether the point is included or discarded.

Every reference point **x** is subjected to a feasibility test before it is evaluated (i.e., before the simulation model is run to determine the value of $f(\mathbf{x})$). The feasibility test consists of checking (one by one) whether the linear constraints imposed by the user are satisfied. An infeasible point **x** is made feasible by formulating and solving a linear programming (LP) problem. The LP (or mixed-integer program, when **x** contains integer variables) has the goal of finding a feasible $\mathbf{x}^*$ that minimises the absolute deviation between **x** and $\mathbf{x}^*$.

The population size is automatically adjusted by the system considering the time that is required to complete one evaluation of $f(\mathbf{x})$ and any time limit on the the system to search. Once the population is generated, the procedure iterates in search of improved outcomes. At each iteration two reference points are selected to create four offspring. Let the parent-reference points be $\mathbf{x}_1$ and $\mathbf{x}_2$, then the offspring $\mathbf{x}_3$ to $\mathbf{x}_6$ are found as follows:

$$\mathbf{x}_3 = \mathbf{x}_1 + d$$
$$\mathbf{x}_4 = \mathbf{x}_1 - d$$

$$\mathbf{x}_5 = \mathbf{x}_2 + d$$
$$\mathbf{x}_6 = \mathbf{x}_2 - d$$

where $d = (\mathbf{x}_1 - \mathbf{x}_2)/3$.

The selection of $\mathbf{x}_1$ and $\mathbf{x}_2$ is biased by the values $f(\mathbf{x}_1)$ and $f(\mathbf{x}_2)$ as well as the tabu search memory functions. An iteration ends by replacing the worst parent with the best offspring, and giving the surviving parent a *tabu-active status* for given number of iterations. In subsequent iterations, the use of two tabu-active parents is forbidden.

## 3.1 Restarting Strategy

In the course of searching for a global optimum, the population may contain many reference points with similar characteristics. That is, in the process of generating offspring from a mixture of high-quality reference points and ordinary reference points, the diversity of the population may tend to decrease. A strategy that remedies this situation considers the creation of new population.

A *restarting* mechanism has the goal of creating a population that is a blend of high-quality points found in earlier explorations (called *elite points*) complemented with points generated in the same way as during the initialisation phase. The restarting procedure, therefore, injects diversity through newly-generated points and preserves quality through the inclusion of elite points.

## 3.2 Adaptive Memory and the Age Strategy

Some of the points in the initial population may have poor objective function values, They may therefore never be chosen to play the role of a parent and would remain in the population until restarting. To diversify the search further, one can increase the *attractiveness* of these unused points over time. This is done by using a form of long-term memory that is different from the conventional frequency-based implementation.

In particular, the notion of *age* is introduced and a measure of *attractiveness* based on the age and the objective function value of a particular point is defined. The idea is to use search history to make reference points not used as parents attractive, by modifying their objective function values according to their age.

At the start of the search process, all the reference points **x** in a population of size *p* have zero age. At the end of the first iteration, there will be *p*-1 reference points from the original population and one new offspring. The ages of the *p*-1 reference points are set to one and that of the new offspring zero. The process then repeats for the subsequent iterations, and the age of every reference point increases by one in each iteration except for the age of the new population member whose age is initialised to zero. (A variant of the this procedure sets the surviving parent's age also to 0.)
Each reference point in the population has an associated age and an objective function value. These two values are used to define a function of attractiveness that makes an old, high-quality point the most attractive. Low-quality points become more attractive as their age increases.

## 3.3 Neural Network Accelerator

The concept behind embedding a neural network is to screen out values **x** that are likely to result in a very poor value of $f(\mathbf{x})$. The neural network is a prediction model that helps the system accelerate the search by avoiding simulation runs whose results can be predicted as inferior. When a neural network is used, information is collected about the objective function values obtained by different optimisation variable settings. This information is then used to train the neural network during the search.

The system automatically determines how much data is needed and how much training should be done, based once again on both the time to perform a simulation and the optimisation time limit provided.

The neural network is trained on the historical data collected during the search and an *error value* is calculated during each training round. This error refers to the accuracy of the network as a prediction model. That is, if the network is used to predict $f(\mathbf{x})$ for

**x**-values found during the search, then the error indicates how good those predictions are. The error term can be calculated by computing the differences between the known $f(\mathbf{x})$ and the predicted objective function values. The training continues until the error reaches a minimum prespecified value.

A neural network accelerator can be used at several risk levels. The risk is associated with the probability of discarding **x** when $f(\mathbf{x})$ is better than $f(\mathbf{x}_{best})$, where $\mathbf{x}_{best}$ is the best solution found so far. The risk level is defined by the number of standard deviations used to determine how close a predicted value is of the best value $f(\mathbf{x}_{best})$. A *risk-averse* user would, for instance, would only discard **x** if is at least three standard deviations larger than $f(\mathbf{x}_{best})$, in a minimisation problem.

# Appendix III: Top 100 share data

Data for the universe of the top 100 stocks is shown in Table 15, ranked by market capitalisation.

*Table 15: Top 100 share data*

| | | | | Top 100 share data | | | | 31/5/00 |
|---|---|---|---|---|---|---|---|---|
| | Share Code | Share | Forecast 2-year total return | β | Market capitalisation | Trade | Forecast variance $\sigma_{e_i}$ | Regression points | Limited history warning |
| | | | (%/p.a) | (x) | (Rm) | (Rm/mth) | (frac) | (no.) | |
| 1 | AGL | Anglo | 52 | 1.24 | 117157 | 4119 | 0.100 | 36 | |
| 2 | RCH | Richemont | 28 | 0.68 | 86652 | 1701 | 0.076 | 36 | |
| 3 | BIL | Billiton | 63 | 1.15 | 55589 | 1245 | 0.104 | 35 | |
| 4 | DBR | De Beers | 61 | 0.97 | 55477 | 1981 | 0.092 | 36 | |
| 5 | OML | Old Mutual | 33 | 1.26 | 52531 | 1209 | 0.040 | 11 | ! |
| 6 | MCE | MCell | 7 | 1.49 | 45343 | 563 | 0.113 | 36 | |
| 7 | FSR | FirstRand | 37 | 1.28 | 43562 | 840 | 0.102 | 36 | |
| 8 | DDT | Didata | 52 | 0.90 | 43091 | 1809 | 0.118 | 36 | |
| 9 | SAB | SAB | 34 | 0.98 | 38684 | 1119 | 0.061 | 36 | |
| 10 | AMS | Amplats | 37 | 0.48 | 35461 | 961 | 0.095 | 36 | |
| 11 | SBC | SBIC | 52 | 1.21 | 34387 | 831 | 0.098 | 36 | |
| 12 | NED | Nedcor | 52 | 0.99 | 30651 | 723 | 0.075 | 36 | |
| 13 | RMT | Rembrandt | 68 | 0.96 | 27457 | 691 | 0.078 | 36 | |
| 14 | SOL | Sasol | 55 | 1.21 | 24149 | 902 | 0.115 | 36 | |
| 15 | SLM | Sanlam | 47 | 1.03 | 21369 | 885 | 0.086 | 19 | ! |
| 16 | INT | Investec | 42 | 1.00 | 20208 | 408 | 0.076 | 36 | |
| 17 | LLA | Liberty | 61 | 1.05 | 17160 | 240 | 0.076 | 9 | ! |
| 18 | JNC | Johnnic | 47 | 1.32 | 16227 | 505 | 0.132 | 36 | |
| 19 | BOE | BOE | 66 | 1.24 | 15610 | 379 | 0.110 | 36 | |
| 20 | ASA | Absa | 55 | 1.50 | 15565 | 608 | 0.100 | 36 | |
| 21 | BVT | Bidvest | 37 | 0.92 | 14474 | 446 | 0.063 | 36 | |
| 22 | IMP | Implats | 57 | 0.62 | 14169 | 566 | 0.130 | 36 | |
| 23 | IPL | Imperial | 49 | 0.98 | 12140 | 411 | 0.069 | 36 | |
| 24 | TBS | Tigbrands | 3 | 0.88 | 11391 | 394 | 0.076 | 36 | |
| 25 | SAP | Sappi | 63 | 1.61 | 11356 | 759 | 0.148 | 36 | |
| 26 | LON | Lonmin | 9 | 0.47 | 11208 | 75 | 0.095 | 25 | |
| 27 | RMH | RMBH | 46 | 1.31 | 10370 | 167 | 0.133 | 36 | |
| 28 | BAR | Barlows | 34 | 1.21 | 8958 | 478 | 0.093 | 36 | |
| 29 | GSC | Gensec | 60 | 1.30 | 8344 | 228 | 0.115 | 36 | |
| 30 | ABL | Abil | 66 | 1.30 | 8294 | 327 | 0.177 | 33 | |
| 31 | NPK | Nampak | 35 | 1.11 | 8042 | 232 | 0.128 | 36 | |
| 32 | FDS | Fedsure | 16 | 1.39 | 6418 | 245 | 0.101 | 36 | |
| 33 | MTC | Metcash | 50 | 1.26 | 6370 | 283 | 0.083 | 36 | |
| 34 | ABI | ABI | 48 | 0.82 | 6009 | 65 | 0.102 | 36 | |
| 35 | DTC | Datatec | 119 | 1.32 | 5870 | 621 | 0.186 | 36 | |
| 36 | PIK | Pick 'n Pay | 23 | 1.28 | 5295 | 63 | 0.154 | 36 | |
| 37 | MET | Metlife | 75 | 1.15 | 5283 | 165 | 0.117 | 36 | |
| 38 | PEP | Pepkor | 76 | 1.31 | 5171 | 162 | 0.163 | 36 | |
| 39 | AFB | Forbes | 36 | 1.23 | 5147 | 121 | 0.081 | 36 | |
| 40 | JDG | JD Group | 44 | 1.40 | 5093 | 219 | 0.127 | 36 | |
| 41 | AIN | Avmin | 41 | 1.81 | 5070 | 181 | 0.108 | 20 | |
| 42 | SHF | Steinhoff | 33 | 1.04 | 4692 | 69 | 0.100 | 21 | |
| 43 | PON | Profurn | 58 | 1.42 | 4552 | 244 | 0.129 | 36 | |
| 44 | CAS | Cadschweppes | 34 | 0.57 | 4383 | 35 | 0.071 | 36 | |
| 45 | CRH | Corohold | 51 | 1.44 | 3882 | 65 | 0.083 | 36 | |

*Table 16: Top 100 share data (continued)*

| | | | | | | | | | |
|---|---|---|---|---|---|---|---|---|---|
| 46 | ISC | Iscor | 85 | 0.40 | 3871 | 237 | 0.153 | 36 | |
| 47 | AFI | Aflife | 44 | 1.10 | 3808 | 39 | 0.135 | 36 | |
| 48 | AFX | Afrox | 30 | 0.78 | 3742 | 38 | 0.110 | 36 | |
| 49 | ECO | Edcon | 40 | 1.02 | 3707 | 163 | 0.187 | 36 | |
| 50 | SHP | Shoprite | 33 | 0.82 | 3699 | 44 | 0.111 | 36 | |
| 51 | TNT | Tongaat | 20 | 0.85 | 3407 | 118 | 0.125 | 36 | |
| 52 | CPX | Comparex | 118 | 1.02 | 3361 | - | 0.150 | 3 | ! |
| 53 | MAF | M & F | 21 | 1.03 | 3262 | 26 | 0.099 | 36 | |
| 54 | NCL | New Clicks | 27 | 1.05 | 2884 | 94 | 0.097 | 36 | |
| 55 | FOS | Foschini | 37 | 1.28 | 2867 | 87 | 0.108 | 36 | |
| 56 | WHL | Woolworths | 51 | 0.72 | 2865 | 103 | 0.138 | 32 | |
| 57 | SNT | Santam | 57 | 1.03 | 2865 | 41 | 0.082 | 36 | |
| 58 | RBH | Rebhold | 66 | 1.25 | 2854 | 128 | 0.138 | 36 | |
| 59 | SPG | Super Group | 41 | 1.36 | 2714 | 106 | 0.143 | 36 | |
| 60 | SFT | Softline | 71 | 1.16 | 2509 | 164 | 0.180 | 36 | |
| 61 | AEG | Aveng | 35 | 0.94 | 2386 | 88 | 0.115 | 11 | ! |
| 62 | PPC | PP Cement | 36 | 0.98 | 2299 | 25 | 0.119 | 36 | |
| 63 | RAH | RAHold | 68 | 1.04 | 2285 | 50 | 0.154 | 36 | |
| 64 | XCH | Ixchange | 116 | 1.28 | 2254 | 187 | 0.372 | 33 | |
| 65 | AHV | Afharv | 81 | 1.21 | 2241 | 61 | 0.207 | 33 | |
| 66 | TRU | Truworths | 34 | 1.45 | 2239 | 67 | 0.212 | 25 | |
| 67 | AVI | AVI | 43 | 0.86 | 2178 | 52 | 0.104 | 11 | ! |
| 68 | KER | Kersaf | 64 | 0.76 | 2132 | 80 | 0.115 | 36 | |
| 69 | AFE | AECI | 63 | 0.83 | 2088 | 92 | 0.172 | 36 | |
| 70 | RLO | Reunert | 51 | 1.20 | 1897 | 86 | 0.118 | 36 | |
| 71 | CPT | Captall | 38 | 1.58 | 1881 | 82 | 0.142 | 36 | |
| 72 | ELH | Ellerines | 29 | 1.55 | 1845 | 110 | 0.106 | 36 | |
| 73 | WLO | Wooltru | 38 | 1.10 | 1754 | 52 | 0.139 | 36 | |
| 74 | BAT | Brait | 50 | 1.25 | 1720 | 82 | 0.192 | 36 | |
| 75 | ILV | Illovo | 31 | 0.64 | 1666 | 84 | 0.100 | 36 | |
| 76 | MLB | Malbak | 57 | 0.84 | 1644 | 47 | 0.115 | 36 | |
| 77 | ALT | Altech | 58 | 0.90 | 1504 | 29 | 0.149 | 36 | |
| 78 | UTR | Unitrans | 62 | 0.78 | 1405 | 29 | 0.119 | 36 | |
| 79 | EDC | Educor | 112 | 1.10 | 1355 | 89 | 0.142 | 36 | |
| 80 | AVS | Avis | 60 | 0.87 | 1336 | 48 | 0.095 | 36 | |
| 81 | NHM | Northam | 62 | 0.26 | 1312 | 28 | 0.193 | 36 | |
| 82 | OTK | OTK | 53 | 0.93 | 1282 | 68 | 0.095 | 36 | |
| 83 | MUR | M & R | 50 | 1.67 | 1211 | 49 | 0.156 | 36 | |
| 84 | NTC | Netcare | 57 | 1.49 | 1188 | 30 | 0.167 | 36 | |
| 85 | PGR | Peregrine | 91 | 1.30 | 1138 | 57 | 0.152 | 24 | |
| 86 | UNF | Unifer | 54 | 1.24 | 1112 | 40 | 0.136 | 14 | ! |
| 87 | AMB | AMB | 64 | 1.00 | 1029 | 74 | 0.189 | 31 | |
| 88 | POW | Powertech | 21 | 0.70 | 1016 | 26 | 0.134 | 36 | |
| 89 | TIW | Tiger Wheels | 83 | 0.91 | 840 | 42 | 0.118 | 36 | |
| 90 | OUS | Outsors | 160 | 1.61 | 831 | 252 | 0.482 | 32 | |
| 91 | PRI | Primedia | 53 | 1.49 | 783 | 51 | 0.121 | 36 | |
| 92 | LST | Leisurenet | 65 | 1.13 | 664 | 41 | 0.147 | 36 | |
| 93 | CCH | CCH | 125 | 1.30 | 592 | 243 | 0.292 | 33 | |
| 94 | TRT | Tourvest | 100 | 0.94 | 520 | 35 | 0.144 | 36 | |
| 95 | RAD | RAD | 45 | 1.28 | 519 | 26 | 0.236 | 24 | |
| 96 | CRS | Carson | 65 | 1.25 | 448 | 28 | 0.229 | 36 | |
| 97 | SPS | Spescom | 139 | 1.53 | 263 | 24 | 0.155 | 36 | |
| 98 | USK | Usko | 7 | 1.51 | 154 | 92 | 0.316 | 36 | |
| 99 | SPI | Spicer | 55 | 1.78 | 56 | 28 | 1.168 | 36 | |
| 100 | OMC | Omnicor | 28 | 0.86 | - | 81 | 0.140 | 36 | |
| Arithmetic mean | | | 53 | 1.11 | 11086 | 328 | 0.144 | - | - |
| - | CI01 | Alsi | 46 | 1.00 | - | - | - | - | - |

# Appendix IV: Comparison of heuristic methods

| Cardinality-unconstrained efficient frontier | | | | | | | | | |
|---|---|---|---|---|---|---|---|---|---|
| **Mixed-integer solver** | | | | | | | | | |
| w | Return | Risk | Objective function | No. of stocks | | | | | |
|  | (%) | (%) | (%) | (No.) | | | | | |
| 0.0000 | 81.8 | 0.366 | 81.767 | 100 | | | | | |
| 0.9930 | 81.2 | 0.353 | 0.218 | 100 | | | | | |
| 0.9940 | 79.2 | 0.340 | 0.137 | 100 | | | | | |
| 0.9945 | 75.9 | 0.322 | 0.098 | 100 | | | | | |
| 0.9950 | 73.5 | 0.308 | 0.061 | 100 | | | | | |
| 0.9955 | 69.8 | 0.291 | 0.024 | 100 | | | | | |
| 0.9957 | 65.7 | 0.273 | 0.010 | 100 | | | | | |
| 0.9960 | 61.8 | 0.257 | -0.009 | 100 | | | | | |
| 0.9970 | 57.7 | 0.242 | -0.068 | 100 | | | | | |
| 0.9980 | 54.8 | 0.236 | -0.126 | 100 | | | | | |
| 0.9981 | 50.9 | 0.228 | -0.131 | 100 | | | | | |
| 0.9982 | 47.7 | 0.222 | -0.136 | 100 | | | | | |
| 0.9990 | 43.1 | 0.214 | -0.171 | 100 | | | | | |
| 1.0000 | 37.7 | 0.211 | -0.211 | 100 | | | | | |
| **TS frontier** | | | | | **TS efficiency** | | | | |
| w | Return | Risk | Objective function | No. of stocks | TS absolute error | | | Solution time | Best trial | Total trials |
|  | (%) | (%) | (%) | (No.) | Return (%) | Risk (%) | Objective (%) | (min) | (no.) | (no.) |
| 0.0000 | 77.7 | 0.361 | 77.713 | 100 | 4.96 | 1.42 | 4.96 | 43 | 550 | 1000 |
| 0.9930 | 77.2 | 0.350 | 0.193 | 100 | 4.91 | 0.89 | 11.41 | 70 | 161 | 427 |
| 0.9940 | 76.1 | 0.339 | 0.120 | 100 | 3.87 | 0.52 | 12.17 | 60 | 67 | 577 |
| 0.9945 | 72.9 | 0.320 | 0.083 | 100 | 3.99 | 0.56 | 15.22 | 60 | 410 | 741 |
| 0.9950 | 71.0 | 0.307 | 0.049 | 100 | 3.40 | 0.34 | 18.87 | 60 | 601 | 680 |
| 0.9955 | 67.6 | 0.290 | 0.015 | 100 | 3.20 | 0.44 | 36.40 | 47 | 66 | 436 |
| 0.9957 | 64.4 | 0.274 | 0.004 | 100 | 1.93 | 0.35 | 60.94 | 60 | 100 | 636 |
| 0.9960 | 61.0 | 0.259 | -0.014 | 100 | 1.29 | 0.83 | 61.54 | 60 | 54 | 675 |
| 0.9970 | 57.0 | 0.244 | -0.072 | 100 | 1.10 | 0.83 | 5.71 | 60 | 205 | 848 |
| 0.9980 | 54.5 | 0.237 | -0.128 | 100 | 0.49 | 0.44 | 1.25 | 60 | 877 | 970 |
| 0.9981 | 50.8 | 0.228 | -0.131 | 100 | 0.05 | 0.04 | 0.11 | 60 | 331 | 932 |
| 0.9982 | 48.2 | 0.223 | -0.136 | 100 | 1.04 | 0.52 | 0.18 | 60 | 435 | 807 |
| 0.9990 | 39.9 | 0.211 | -0.171 | 100 | 7.28 | 1.29 | 0.22 | 60 | 756 | 869 |
| 1.0000 | 35.7 | 0.215 | -0.215 | 100 | 5.30 | 1.77 | 1.77 | 60 | 694 | 705 |
|  | **Median** | | | | **3.30** | **0.54** | **8.56** | **60** | **371** | **723** |
|  | **Standard deviation** | | | | **2.06** | **0.46** | **20.62** | **6** | **271** | **175** |
|  | **Mean** | | | | **3.06** | **0.73** | **16.48** | **59** | **379** | **736** |
|  | **Combined mean** | | | | **1.89** | | **-** | **-** | **-** | **-** |
| **GA frontier** | | | | | **GA efficiency** | | | | |
| w | Return | Risk | Objective function | No. of stocks | GA absolute error | | | Solution time | Best trial | Total trials |
|  | (%) | (%) | (%) | (No.) | Return (%) | Risk (%) | Objective (%) | (min) | (no.) | (no.) |
| 0.0000 | 81.77 | 0.366 | 81.767 | 100 | 0.00 | 0.00 | 0.00 | 16 | 13752 | 13752 |
| 0.9930 | 81.25 | 0.354 | 0.217 | 100 | 0.09 | 0.20 | 0.08 | 25 | 39486 | 39486 |
| 0.9940 | 78.96 | 0.339 | 0.137 | 100 | 0.25 | 0.35 | 0.01 | 26 | 17055 | 17055 |
| 0.9945 | 74.61 | 0.315 | 0.097 | 100 | 1.72 | 2.22 | 0.11 | 7 | 8320 | 8320 |
| 0.9950 | 73.63 | 0.309 | 0.061 | 100 | 0.18 | 0.22 | 0.01 | 13 | 27146 | 27146 |
| 0.9955 | 69.85 | 0.292 | 0.024 | 100 | 0.02 | 0.03 | 0.14 | 7 | 15885 | 15885 |
| 0.9957 | 65.15 | 0.271 | 0.010 | 100 | 0.86 | 0.88 | 0.38 | 7 | 13333 | 13333 |
| 0.9960 | 61.62 | 0.256 | -0.009 | 100 | 0.28 | 0.27 | 0.14 | 20 | 40256 | 40256 |
| 0.9970 | 57.64 | 0.242 | -0.068 | 100 | 0.06 | 0.04 | 0.00 | 9 | 22998 | 22998 |
| 0.9980 | 56.49 | 0.239 | -0.126 | 100 | 3.14 | 1.47 | 0.02 | 5 | 6578 | 6578 |
| 0.9981 | 51.28 | 0.229 | -0.131 | 100 | 0.82 | 0.38 | 0.06 | 3 | 5768 | 5768 |
| 0.9982 | 46.66 | 0.220 | -0.136 | 100 | 2.10 | 0.81 | 0.01 | 7 | 23601 | 23601 |
| 0.9990 | 43.25 | 0.215 | -0.171 | 100 | 0.44 | 0.17 | 0.10 | 4 | 11445 | 11445 |
| 1.0000 | 38.38 | 0.211 | -0.211 | 100 | 1.92 | 0.08 | 0.08 | 9 | 12238 | 12238 |
|  | **Median** | | | | **0.36** | **0.24** | **0.07** | **8** | **14819** | **14819** |
|  | **Standard deviation** | | | | **0.95** | **0.62** | **0.09** | **7** | **10656** | **10656** |
|  | **Mean** | | | | **0.85** | **0.51** | **0.08** | **11** | **18419** | **18419** |
|  | **Combined mean** | | | | **0.68** | | **-** | **-** | **-** | **-** |

*Table 17: Heuristic test data*